\documentclass[journal]{IEEEtran}
\usepackage{cite}
\ifCLASSINFOpdf
\else
\fi 
\usepackage{multirow}
\usepackage{flushend}
\usepackage{graphicx}
\usepackage{colortbl}
\graphicspath{{Figures_eeg/}}
\usepackage{caption}
\usepackage{capt-of}
\usepackage{balance}
\usepackage{subfigure}
\usepackage{fixltx2e}
\usepackage{setspace}
\usepackage{amsmath}
\usepackage{hyperref}
\usepackage{array}
\newcommand{\specialcell}[2][c]{%
\begin{tabular}[#1]{@{}c@{}}#2\end{tabular}}
\begin{document}
\title{Source Aware Deep Learning Framework for Hand Kinematic Reconstruction using EEG Signal} \author{Sidharth~Pancholi$^\ast$,~\IEEEmembership{Member,~IEEE,} Amita~Giri$^\ast$,~\IEEEmembership{Student Member,~IEEE,}
Anant~Jain,~\IEEEmembership{Student Member,~IEEE,}\\ Lalan Kumar,~\IEEEmembership{Member,~IEEE, and} Sitikantha Roy,~\IEEEmembership{Member,~IEEE}

\thanks{This work was supported in part by Prime Minister’s Research Fellowship (PMRF), Ministry of Education (MoE), GoI, with project number PLN08/PMRF (to A. Giri) and in part by DRDO - IITD `JATC' with project number RP03830G.}
\thanks{S. Pancholi, A. Giri and A. Jain are with the Department of Electrical Engineering, Indian Institute of Technology, Delhi, India (e-mail: s.pancholi@ieee.org; Amita.Giri@ee.iitd.ac.in; anantjain@ee.iitd.ac.in)}
\thanks{L. Kumar is with the Department of Electrical Engineering and Bharti School of Telecommunication, Indian Institute of Technology, Delhi, India (e-mail: lkumar@ee.iitd.ac.in)}
\thanks{S. Roy is with the Department of Applied Mechanics, Indian Institute of Technology, Delhi, India (e-mail: sroy@am.iitd.ac.in)}
\thanks{$^\ast$S. Pancholi and A. Giri contributed equally to this work.}}

\maketitle

\begin{abstract}
The ability to reconstruct the kinematic parameters of hand movement using non-invasive electroencephalography (EEG) is essential for strength and endurance augmentation using exosuit/exoskeleton. For system development, the conventional classification based brain computer interface (BCI) controls external devices by providing discrete control signals to the actuator. A continuous kinematic reconstruction from EEG signal is better suited for practical BCI applications. The state-of-the-art multi-variable linear regression (mLR) method provides a continuous estimate of hand kinematics, achieving maximum correlation of upto 0.67 between the measured and the estimated hand trajectory. In this work, three novel source aware deep learning models are proposed for motion trajectory prediction (MTP). In particular, multi layer perceptron (MLP), convolutional neural network - long short term memory (CNN-LSTM), and wavelet packet decomposition (WPD) CNN-LSTM are presented. Additional novelty of the work includes utilization of brain source localization (using sLORETA) for the reliable decoding of motor intention. The information is utilized for channel selection and accurate EEG time segment selection. Performance of the proposed models are compared with the traditionally utilised mLR technique on the real grasp and lift (GAL) dataset. Effectiveness of the proposed framework is established using the Pearson correlation coefficient and trajectory analysis. A significant improvement in the correlation coefficient is observed when compared with state-of-the-art mLR model. Our work bridges the gap between the control and the actuator block, enabling real time BCI implementation. 
 \end{abstract}

\section{Introduction}\label{sec:num1}
\begin{figure}[t]
	\centering
	\includegraphics[width=8.5cm,height=5.8cm]{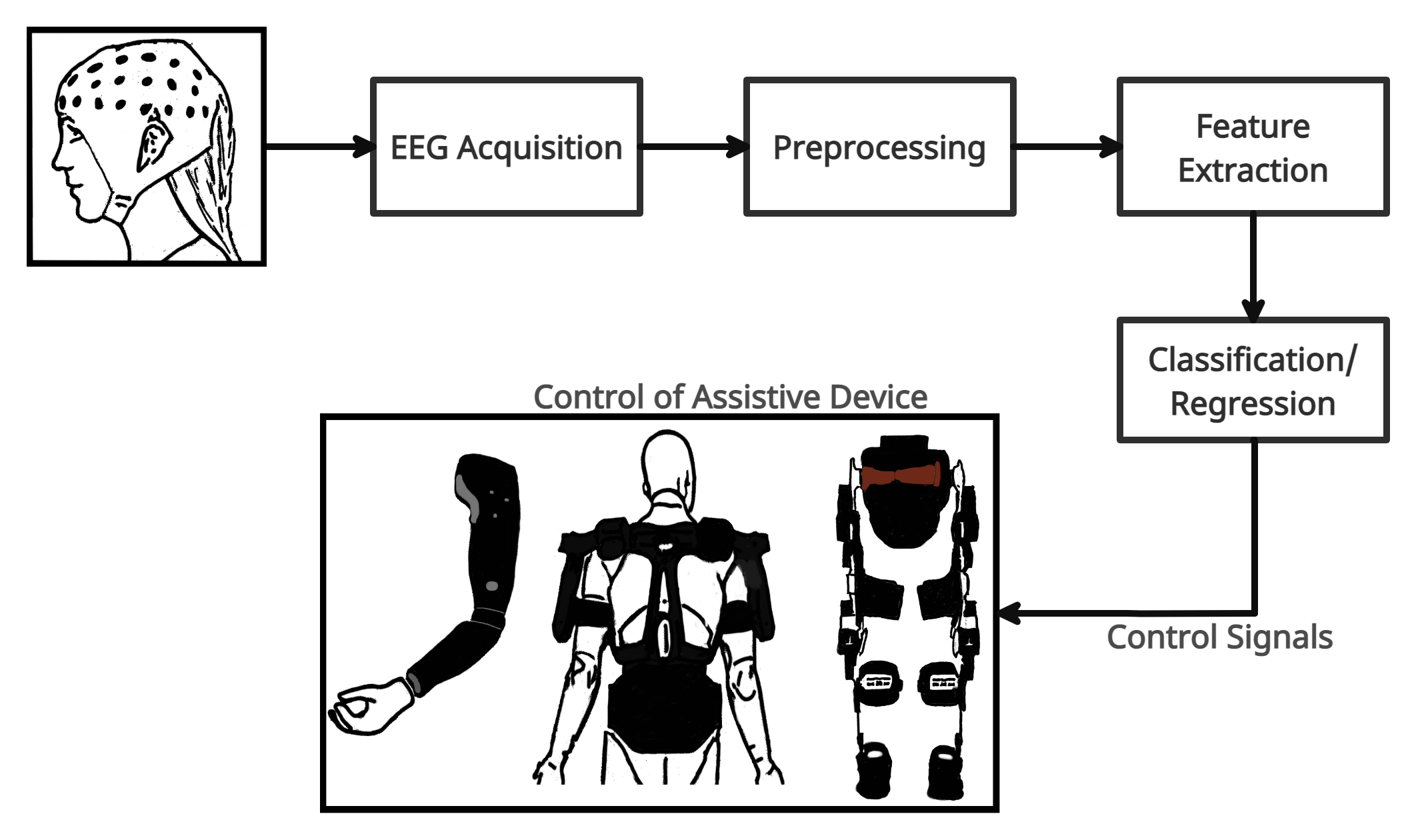}
	\caption{Schematic diagram of EEG signal based BCI system.}
	\label{thematic}
\end{figure}

\IEEEPARstart  Electroencephalography (EEG) signal has been extensively utilized for brain computer interface (BCI) applications because of its high temporal resolution, non-invasive nature, portability, and cost-effectiveness \cite{zhang2019making, kwak2019error, zhang2018temporally, zhang2019eeg, bhagat2016design}. BCI systems facilitates direct connection between the brain and external devices that do not rely on the peripheral nerves and muscles. Hence, BCI based devices like wearable robots \cite{zhang2019eeg,he2015wireless}, exoskeletons \cite{deng2018learning, bhagat2016design,gao2020classification}, and prosthesis \cite{li2019eeg} have gained focus in the recent years. When such external devices are utilized for strength and endurance augmentation, reconstruction of motion trajectory kinematic parameters from EEG signal becomes important. Reliable decoding of motor intentions and accurate timing of the robotic device actuation is fundamental to optimally enhance the subject’s functional improvement. Since, EEG signal has information about the kinematic parameters prior to the actual movement, this time gain along with correct intention mapping will facilitate the real time control of assistive devices. A schematic diagram of EEG signal based BCI systems is shown in the Fig. \ref{thematic}.

Multi-class classification \cite{alazrai2019eeg,he2015common} and regression \cite{robinson2015adaptive} are the two major approaches adopted in EEG based BCI control. Multi-class classification based BCI systems utilize feature extraction and classification method to maximize the inter-class variance and realize decisive planes that separate distinct classes \cite{shajil2020multiclass}. However, regression-based techniques provide more natural control of assistive devices through continuous decoding of EEG signal. Brain activity recorded as EEG signal, exhibits non-stationary nature and therefore requires continuous estimation of optimal and stationary features from the signal. Motion trajectory prediction (MTP) from multi-channel EEG signals was proposed in \cite{robinson2015adaptive} using the multi-variate linear regression (mLR) technique. In particular, Kalman filter based mLR model was utilized to decode 2D hand movement in the horizontal plane. Mean correlation value of 0.60$\pm$0.07 was achieved between the predicted and the measured trajectory. The most adopted hand crafted feature for regression is power spectral density (PSD) from the four frequency bands that include delta (1-3Hz), theta (5-8Hz), alpha (9-12Hz), and beta (14-31Hz) \cite{rangayyan2015biomedical}. In \cite{korik2018decoding}, decoding of 3D imagined hand movement trajectory was studied using PSD based band-power time series (BTS) technique. A significant improvement in accuracy was observed using BTS input when compared with the standard potentials time-series (PTS) input. The most recent work \cite{sosnik2020reconstruction}, demonstrated the feasibility of predicting both actual and imagined 3D trajectories of all arm joints from scalp EEG signals using mLR model. However, the limitation of mLR method is that it demands the linear relationship between the independent (observed event) variables and dependent (predicted event) variable. Furthermore, mLR is extremely sensitive to outliers and poor quality data. If the number of outliers relative to the non-outlier data points is greater than a few, the linear regression model deviates from the true underlying relationship.

In the past few years, deep learning a sub-field of machine learning has achieved breakthrough accuracies in complex and high dimensional data such as image classification \cite{liu2020deep}, emotion recognition \cite{zhang2018spatial}, and machine translation \cite{zhang2015deep}. Deep learning focuses on computational models that typically learn hierarchical representations of the input data through successive non-linear transformations—termed neural networks. In contrast to mLR, the deep learning models are much more robust to outliers and can access very descriptive (non-linear) features that define the underlying relationships fairly well. The convolutional neural network (CNN/ConvNet) has been extensively utilized for BCI applications that include motor imagery (MI) and motor execution (ME) classification \cite{borra2020interpretable}. The widespread use of CNN algorithms in classification application \cite{sun2020automatically,yang2018scene} is due to its capability to extract spatial information from EEG signal. However, it intrinsically disregards the temporal information \cite{amin2019deep}. A very widely known recurrent neural network which effectively utilizes the temporal dependencies based on past information is long short-term memory (LSTM) network \cite{du2020novel}.

\begin{figure}[t]
	\centering
	\includegraphics[width=0.40\textwidth]{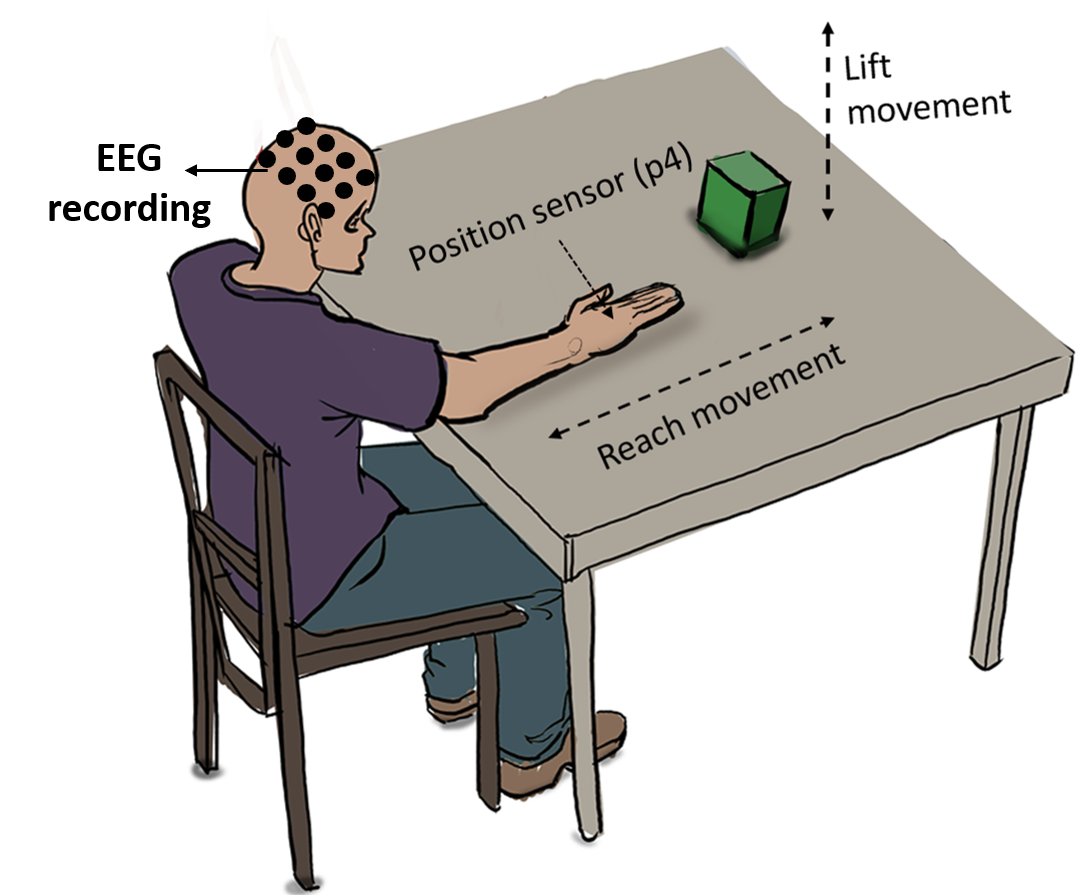}
	\caption{Experimental set-up of reach to grasp movement.}
	\label{exp} 
\end{figure}

In this work, a basic feed forward neural network called multi layer perceptron (MLP) along with CNN-LSTM based hybrid deep learning framework is proposed for hand kinematics prediction. To the best of authors' knowledge, CNN itself has not been utilized for MTP. Reconstruction of hand movement profiles using low frequency EEG have been reported in 2D \cite{presacco2011neural} and 3D spaces \cite{bradberry2010reconstructing}. These results indicate that detailed limb kinematic information could be present in the low frequency components of EEG, and could be decoded using the proposed models. Hence, an advanced version of CNN-LSTM based on wavelet packet decomposition (WPD) is proposed that decompose the EEG signal into sub-bands with increasing resolution towards the lower frequency band \cite{zhang2017classification, sadiq2019motor}. Additional novelty of the work includes utilization of brain source localization (BSL) for the reliable decoding of motor intention. The information is utilized for channel selection and accurate EEG time segment selection. Electrodes placed over the active brain region corresponding to the hand movement are utilized, rather than all the available sensors data for efficient computation. The selected EEG segment is then utilized in the training and testing of the proposed deep learning model. Hence, the proposed framework is called the source aware deep learning framework for hand kinematics prediction.

The rest of the paper is organized as follows. The details of the experimental data and signal pre-processing steps are covered in Section \ref{sec:num2}. The description of existing state-of-the-art model is presented in Section \ref{sec:num3}. The three proposed source aware deep learning models along with the role of brain source localization in MTP is presented in Section \ref{sec:num4}. Performance evaluation metric is reported in Section \ref{sec:num5}. Section \ref{sec:num6} provides a detailed discussion of the results followed by conclusions of our work.

\begin{figure}[t]
	\centering
	\includegraphics[width=0.50\textwidth]{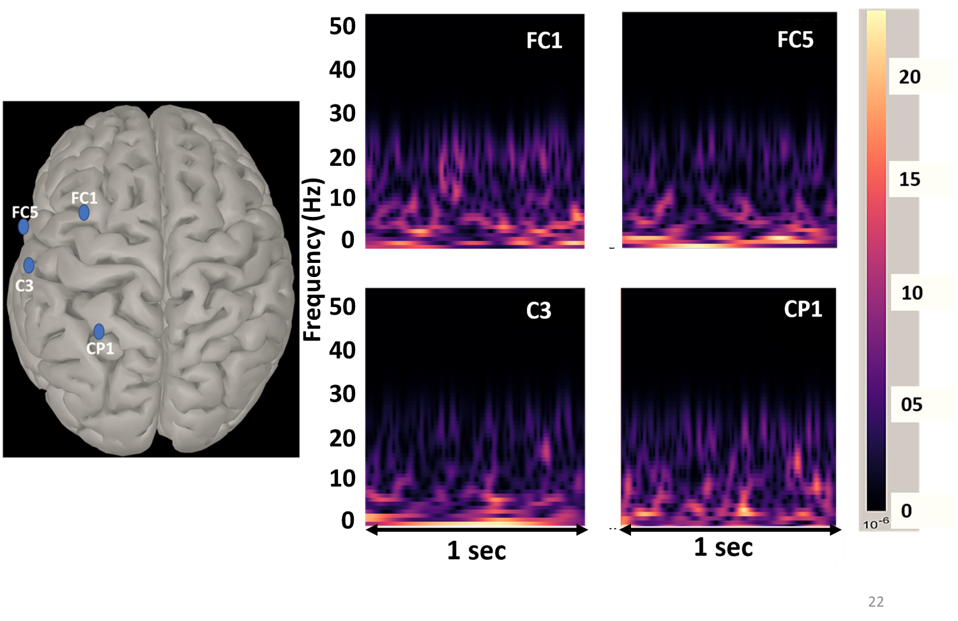}
	\caption{Time-frequency distribution of FC1, FC5, C3, and CP1 EEG channels.}
	\label{wave}
\end{figure}
\section{Experimental Data and Pre-Processing}\label{sec:num2}
In this work, WAY-EEG-GAL (Wearable interfaces for hAnd function recoverY- EEG - grasp and lift) data-set \cite{luciw2014multi} is utilized for MTP. Scalp EEG recordings were collected from twelve healthy subjects for right hand movement. In this experiment, the task to be executed was to reach and grasp the object and lift it stably for a couple of seconds. The participant can then lower the object at its initial position and retract the arm back to resting position. The data acquisition set up for the same is illustrated in Fig. \ref{exp}. A series of such reach to grasp and lift trials were executed for various loads and surface frictions. The beginning of the task and lowering of the object was cued by an LED. The kinematic data was obtained using a position sensor p4 (as shown in Fig. \ref{exp}) normalized between 0 to 1 with initial position as 0 and maximum as 1. This was done to get rid of error due to initial position perturbation. The pre-processing steps followed is detailed next.
\begin{figure*}[t]
	\centering
	\subfigure[]{\includegraphics[width=0.23\textwidth,height=0.20\textwidth]{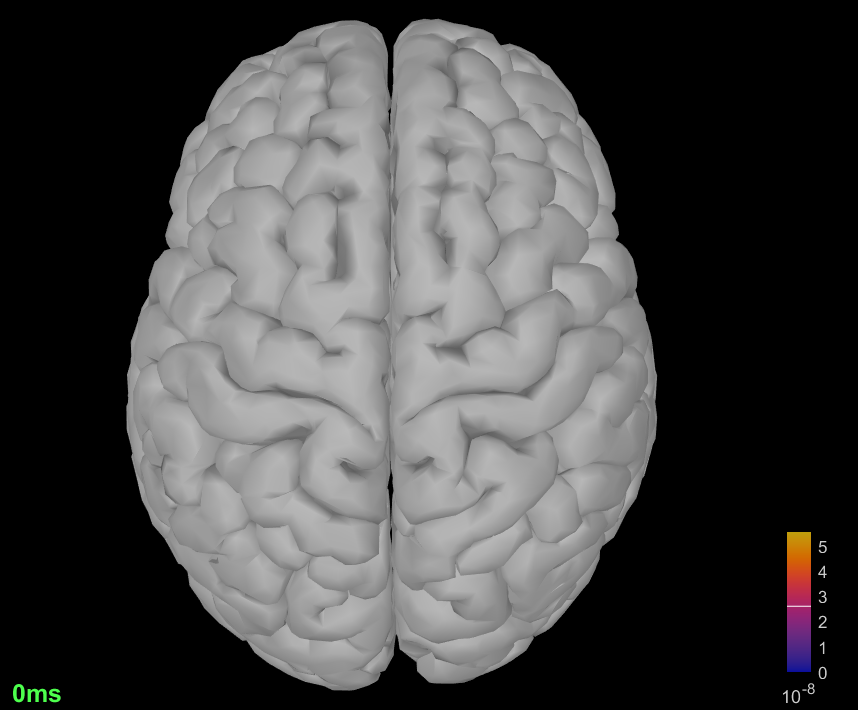}}
	\subfigure[]{\includegraphics[width=0.23\textwidth,height=0.20\textwidth]{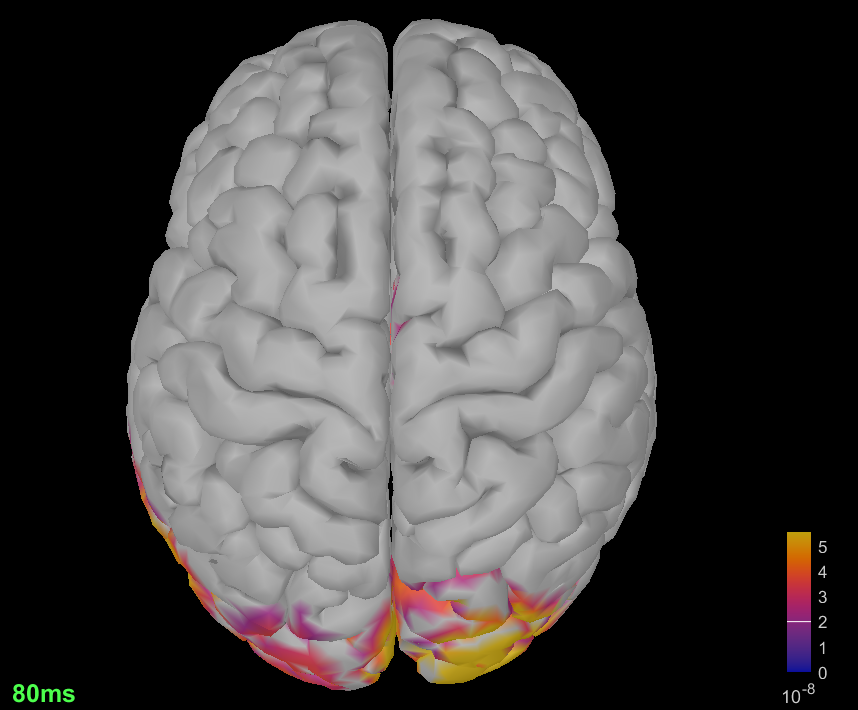}}
	\subfigure[]{\includegraphics[width=0.23\textwidth,height=0.20\textwidth]{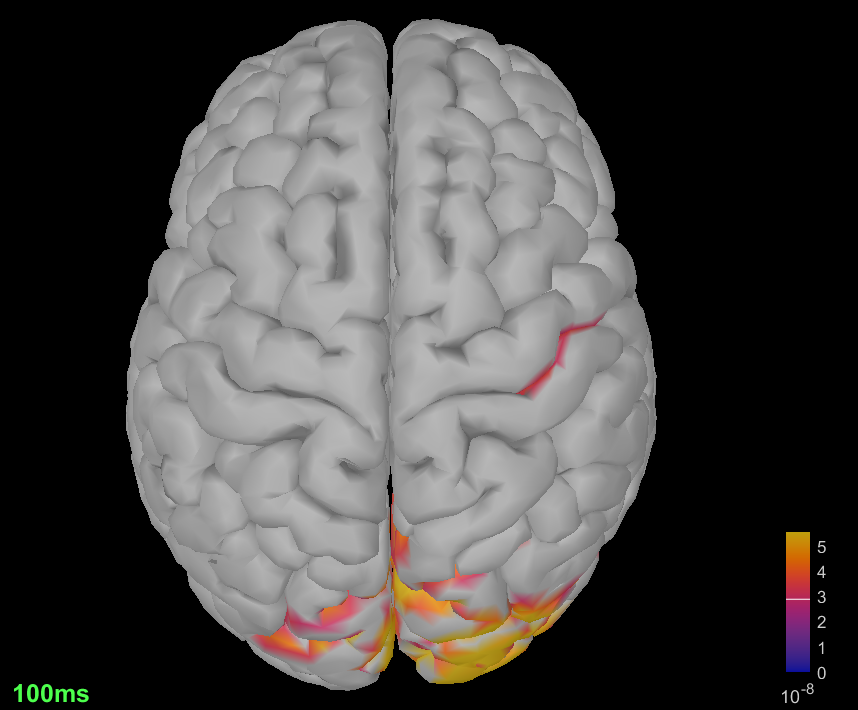}}
	\subfigure[]{\includegraphics[width=0.23\textwidth,height=0.20\textwidth]{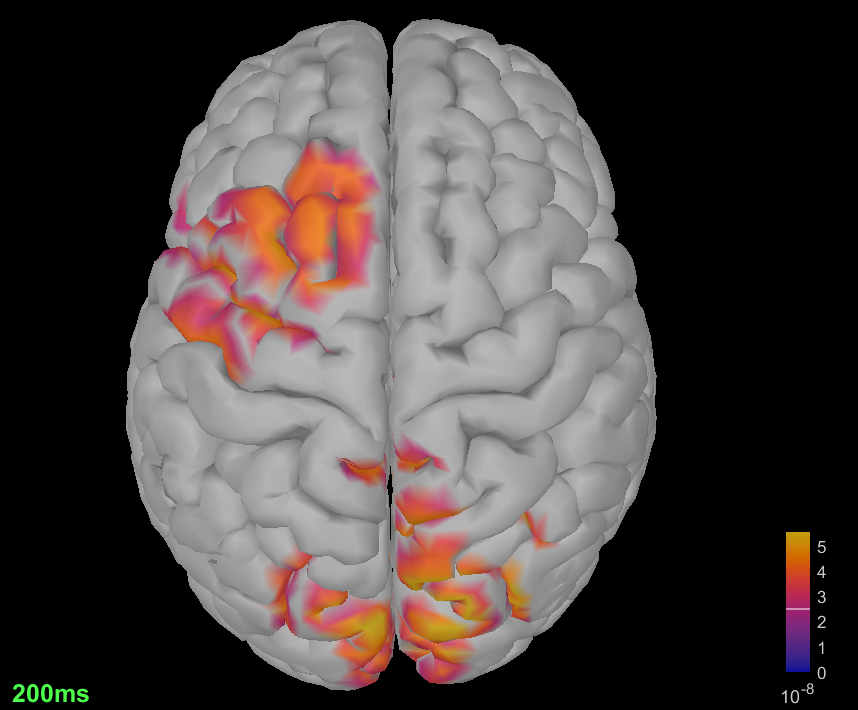}}
	\subfigure[]{\includegraphics[width=0.23\textwidth,height=0.20\textwidth]{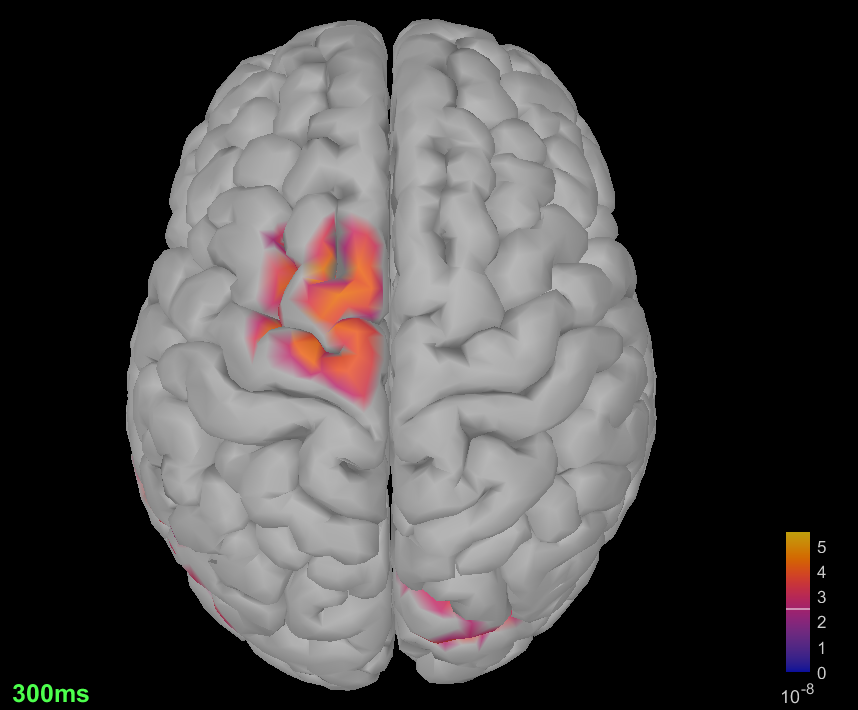}}
	\subfigure[]{\includegraphics[width=0.23\textwidth,height=0.20\textwidth]{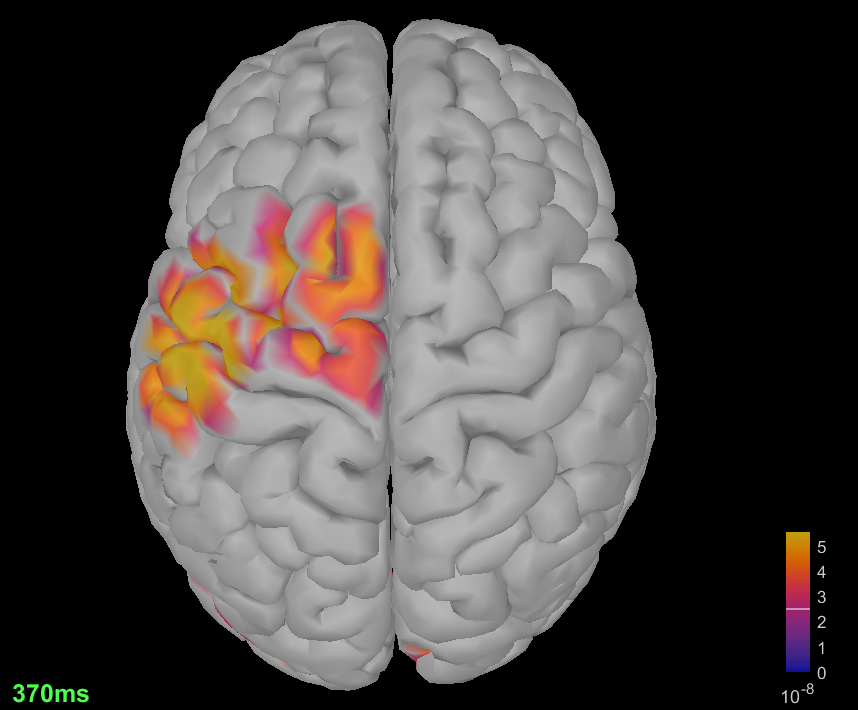}}
	\subfigure[]{\includegraphics[width=0.23\textwidth,height=0.20\textwidth]{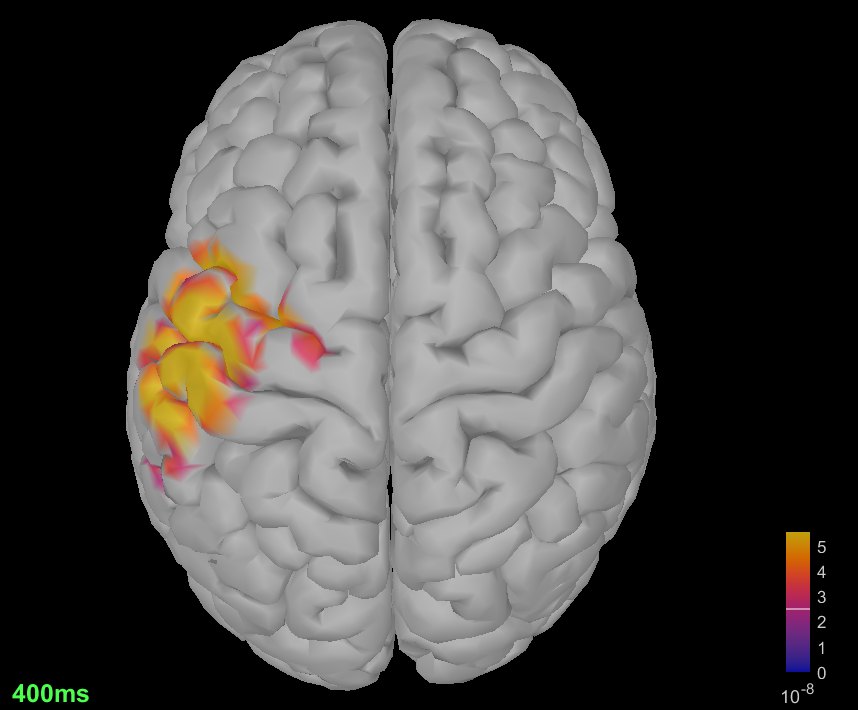}}
	\subfigure[]{\includegraphics[width=0.23\textwidth,height=0.20\textwidth]{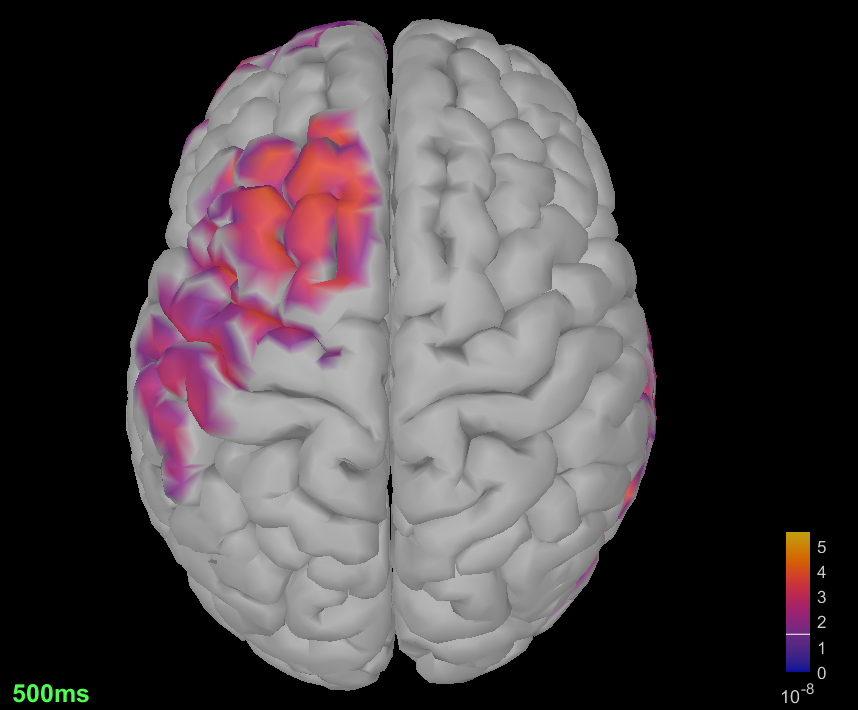}}
	\subfigure[]{\includegraphics[width=0.23\textwidth,height=0.20\textwidth]{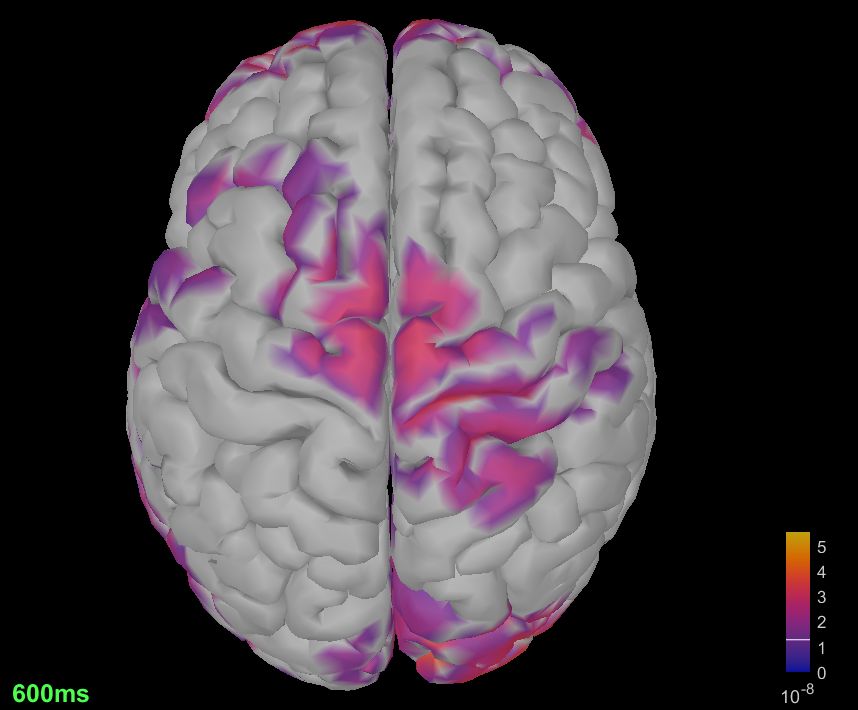}}
	\subfigure[]{\includegraphics[width=0.23\textwidth,height=0.20\textwidth]{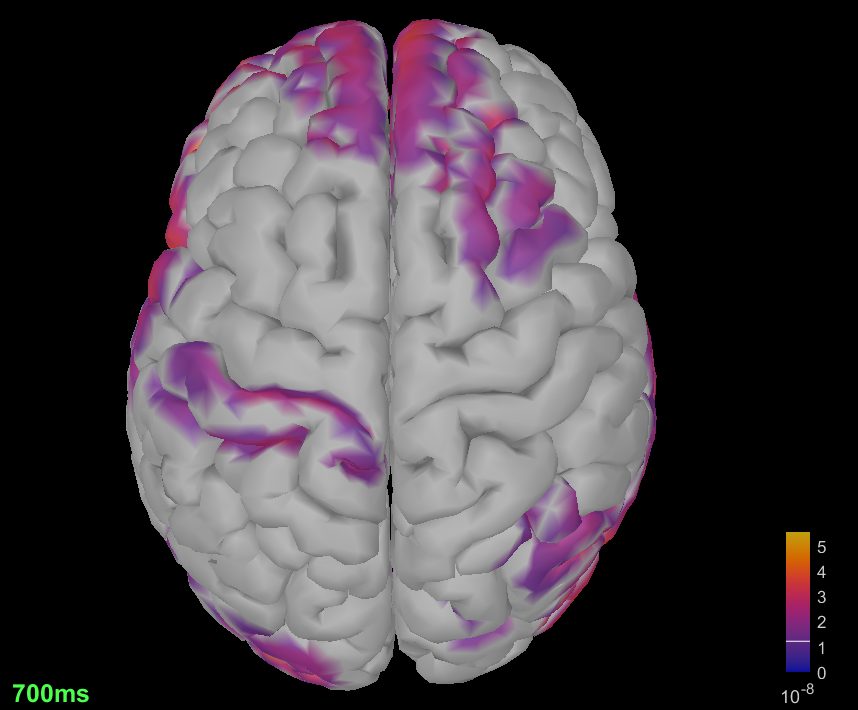}}
	\subfigure[]{\includegraphics[width=0.23\textwidth,height=0.20\textwidth]{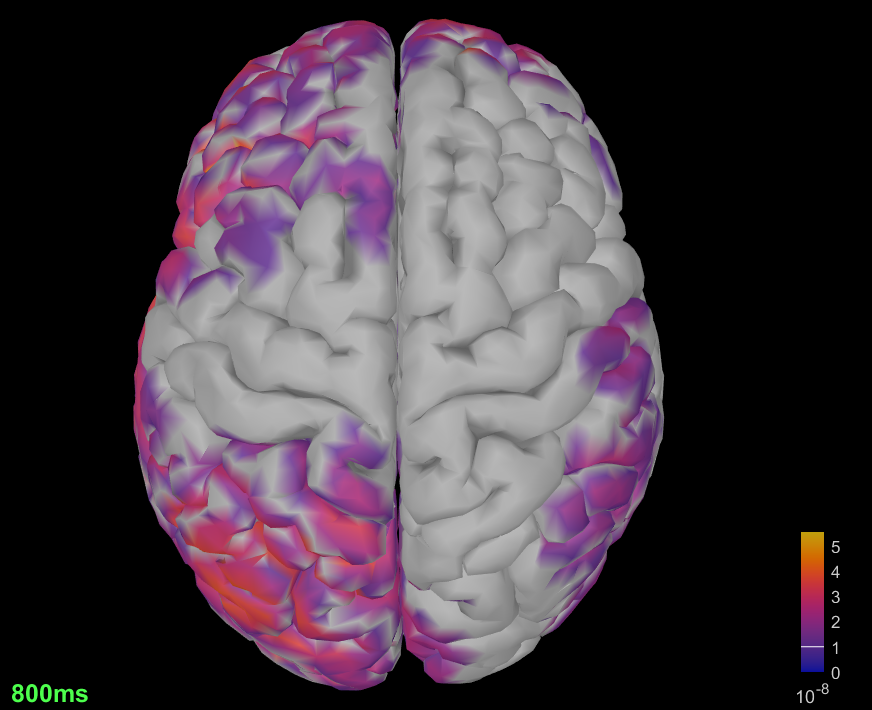}}
	\subfigure[]{\includegraphics[width=0.23\textwidth,height=0.20\textwidth]{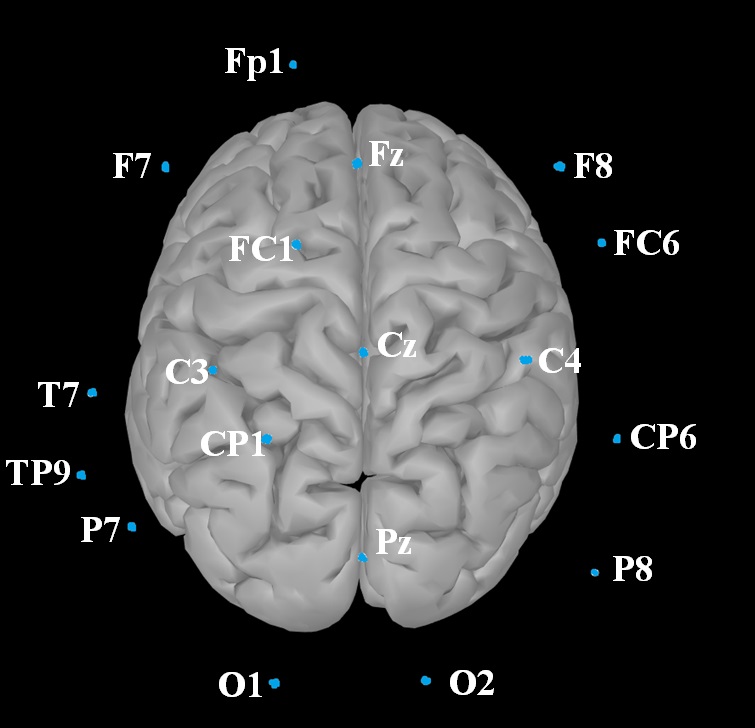}}
	\caption{Brain source localization using sLORETA at different time stamps : (a) 0ms (b) 80ms  (c) 100ms (d) 200ms (e) 300ms (f) 370ms (g) 400ms (h) 500ms (i) 600ms (j) 700ms (k) 800ms. (l) Selected 18 channels of maximum activity using BSL.} 
	\label{result}
	
\end{figure*}

 The kinematics and EEG data were down sampled from 500 Hz to 100 Hz. The time-frequency distribution of EEG signal for a particular subject is shown in the Fig. \ref{wave} for FC1, FC5, C3, and CP1 channel. It may be noted that the maximum power related to right hand movement is present in the delta (0.5-3 Hz), theta (3-7 Hz), and lower alpha (7-12 Hz) range. Hence, the EEG time series was first filtered using zero-phase $4$th order Hamming-windowed sinc FIR (finite impulse response) filter in the range of delta, theta, and lower alpha bands. Subsequently, ICA algorithm was applied to remove artifacts such as eye movement, eye blink, and power line interference. The common average referencing method was used for re-referencing. The EEG signal was finally standardized as
\begin{align}
V_{n}[t]=\frac{v_n[t]-\nu_{n}}{\sigma_{n}}
\label{std}
\end{align}
where $V_n[t]$ is the standardized EEG voltage at time $t$ and at sensor $n$. There are total $N$ number of EEG electrodes. The mean and standard deviation of $v_{n}$ is represented by $\nu_n$ and $\sigma_{n}$ respectively. 

\section{The Existing mLR Model for Kinematic Decoding}\label{sec:num3}
Multi-variate linear regression has been the state-of-the-art technique for BCI based MTP \cite{bradberry2010reconstructing, robinson2015adaptive, sosnik2020reconstruction}. In this section, application of mLR in mapping the EEG time series signal to the kinematic parameters in continuous manner, is briefly detailed. The mLR equations for the mapping are as follows \cite{bradberry2010reconstructing}.
\begin{align}
P_x[t] =& a_x+\sum_{n=1}^{N}\sum_{l=0}^{L}b^{(nl)}_{x}V_n[t-l]\\
P_y[t] =& a_y+\sum_{n=1}^{N}\sum_{l=0}^{L}b^{(nl)}_{y}V_n[t-l]\\
P_z[t] =& a_z+\sum_{n=1}^{N}\sum_{l=0}^{L}b^{(nl)}_{z}V_n[t-l]\\ \nonumber
\end{align}
Here, $P_x[t]$, $P_y[t]$, and $P_z[t]$ are the horizontal, vertical, and depth positions of the hand at time sample $t$, respectively. $V_n[t-l]$ is the standardized voltage at time lag $l$, where the number of time lags is varied from $0$ to $L$. The regression coefficient $a$ and $b$ are estimated by minimizing the loss function during the training phase. In mLR, the multiple independent variables ($V_n[t-l]$) contribute to a dependent kinematic variable ($P_x[t]$, $P_y[t]$, and $P_z[t]$).
\section{Source Aware Deep Learning Models for Hand Kinematic Reconstruction} \label{sec:num4}
In this Section, source aware deep learning models for hand kinematic reconstruction are proposed. In particular, MLP, CNN-LSTM and WPD CNN-LSTM are proposed for the kinematic parameter estimation. As the kinematic movement is embedded in the EEG signature, early detection of intended movement is essential for controlling an external BCI devices for positive real time augmentation. Source localization plays a key role in motor intention mapping. The information is utilized for channel selection and accurate EEG time segment selection. Hence, role of brain source localization on MTP is detailed first followed by the model description. 

\subsection{Role of Brain Source Localization in MTP}
Brain source localization refers to the estimation of active dipole location from noninvasive scalp measurements. It is an ill-posed inverse problem, where the relationship between the EEG scalp potential and neural sources is non-unique, and the solution is highly sensitive to the artifacts. Dipole-fitting and dipole imaging (distributed source model) are two approaches to solve the inverse problem. In the dipole fitting method, small number of active regions are considered in the brain and can be modeled using equivalent dipoles \cite{grech2008review, giri2019head}. The dipole fitting is an over-determined approach to BSL and is solved using a nonlinear optimization technique. This includes subspace-based multiple signal classification (MUSIC) \cite{mosher92_multiple, giri2020brain}, beamforming \cite{Baillet2001}, and genetic \cite{genetic} algorithms. On the other hand, the distributed source model assumes that there are a large number of sources confined in an active region of the brain and is solved using linear optimization techniques, such as minimum norm estimation (MNE) \cite{HAMALAINEN1984}, weighted MNE (WMNE) \cite{hamalainen1994MN}, low resolution electromagnetic tomography (LORETA) \cite{pascual1999loreta}, and standardized low resolution brain electromagnetic tomography (sLORETA) \cite{pascual2002standardized}. EEG signals have been found to be effective for monitoring changes in the human brain state and behavior \cite{lin2020driving}. 

In the present work, sLORETA dipole imaging method is opted for the inverse source localization. In this method, localization inference is performed using images of standardized current density under the constraint of smoothly distributed sources. Source localization plots for the activity under consideration (right hand grasp and lift execution task) are given in Fig.~\ref{result}. There are total 11 images of cortical surface activation sliced temporally. The result shown corresponds to single trial EEG of the subject 3 and is reproducible for the other trials as well. Visual cue for the start of the activity was presented at 0ms. It may be noted that the brain region responsible for visual processing (occipital lobe) shows neural activation after 80-120ms of the visual cue (Fig.~\ref{result}(b)-(c)). Prompted hand movement information was transferred to the sensory motor region at around 200-300ms (Fig.~\ref{result}(d)-(e)). In response to right hand movement, contralateral motor cortex i.e. left motor cortex gets elucidated at 370-400ms (Fig.~\ref{result}(f)-(g)). Motor related neural activity is observed thereafter (Fig.~\ref{result}(h)-(k)). It was observed that the Subject actually performed hand movement at 620-650ms after the cue was shown. Hence, it may be concluded that the EEG source localization can provide the intended hand movement information approximately 350ms prior to the actual hand movement. 
\begin{figure}[t]
	\centering
	\includegraphics[width=8cm,height=7cm]{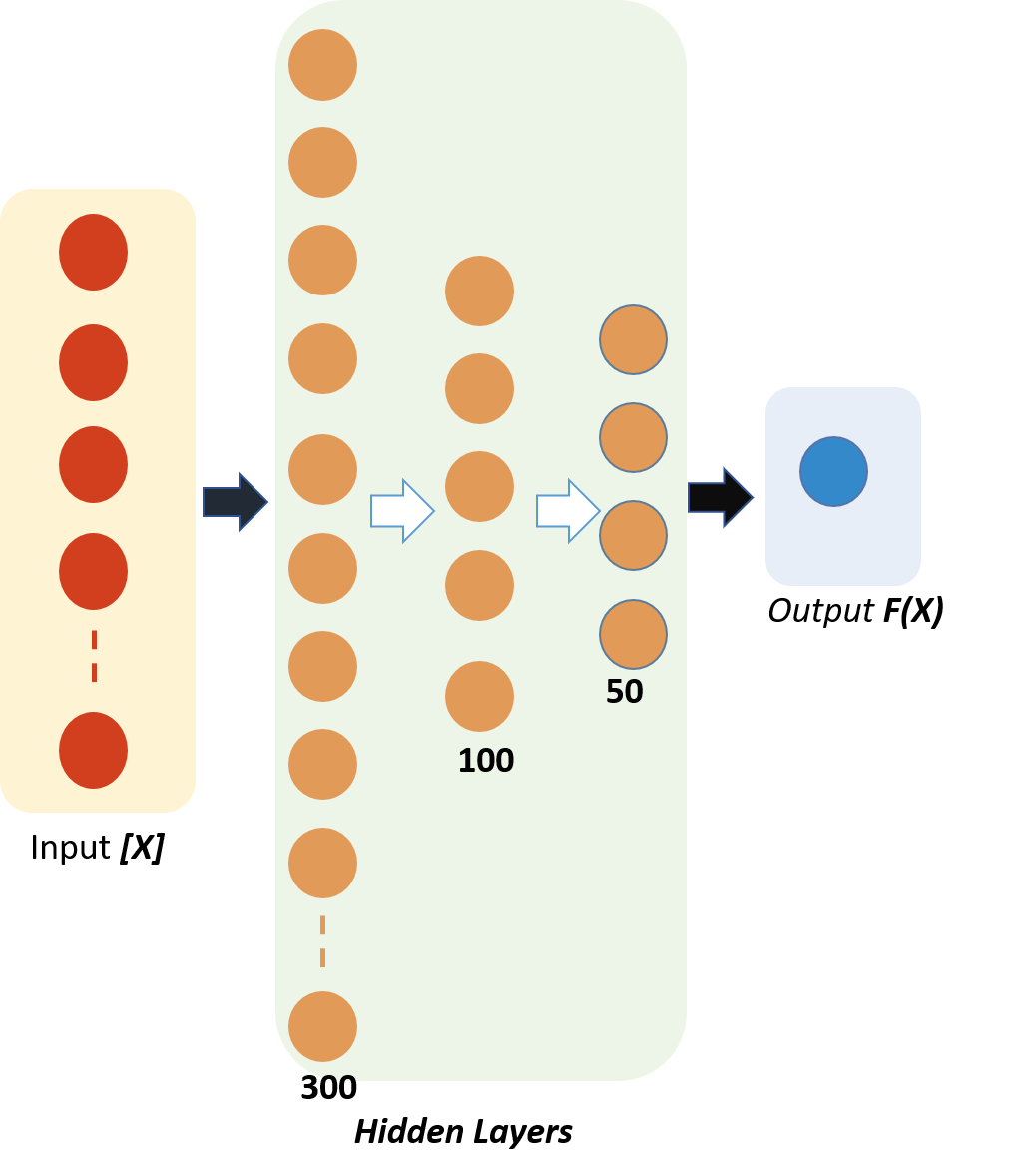}
	\caption{Multi-layer perceptron based regression modeling}
	\label{MLP}
\end{figure}

The transfer delay of visual cue intention to movement onset is 620-650ms including 350ms delay of sensory motor region activation to actual hand movement. This information is utilized to select trials with a delay of up to 700ms between the LED cue and the actual movement. Trials with response times more than 700ms are excluded from this study. In a trial, kinematic data is taken from the movement onset and EEG data was taken starting from the LED cue (0ms). The length of EEG and kinematic data were made equal by removing the EEG samples from the end of a trial. In addition to that, electrodes placed over the maximal neural activation region corresponding to the hand movement were utilized. In particular, electrodes on the left hemisphere (Fp1, F7, FC1, T7, C3, TP9, CP1, P7, O1), near the midline (Fz, Cz, Pz) and on the right hemisphere (F8, FC6, C4, CP6, P8, O2) were utilised as shown in Fig.~\ref{result}(l). Hence, rather than using all the 32 EEG channels for kinematics reconstruction, only 18 channels of maximum activity were chosen. The selected EEG data was utilized in the training and testing of the proposed deep learning model.
\begin{figure*}[t]
	\centering
	\includegraphics[width=17cm,height=9.6cm]{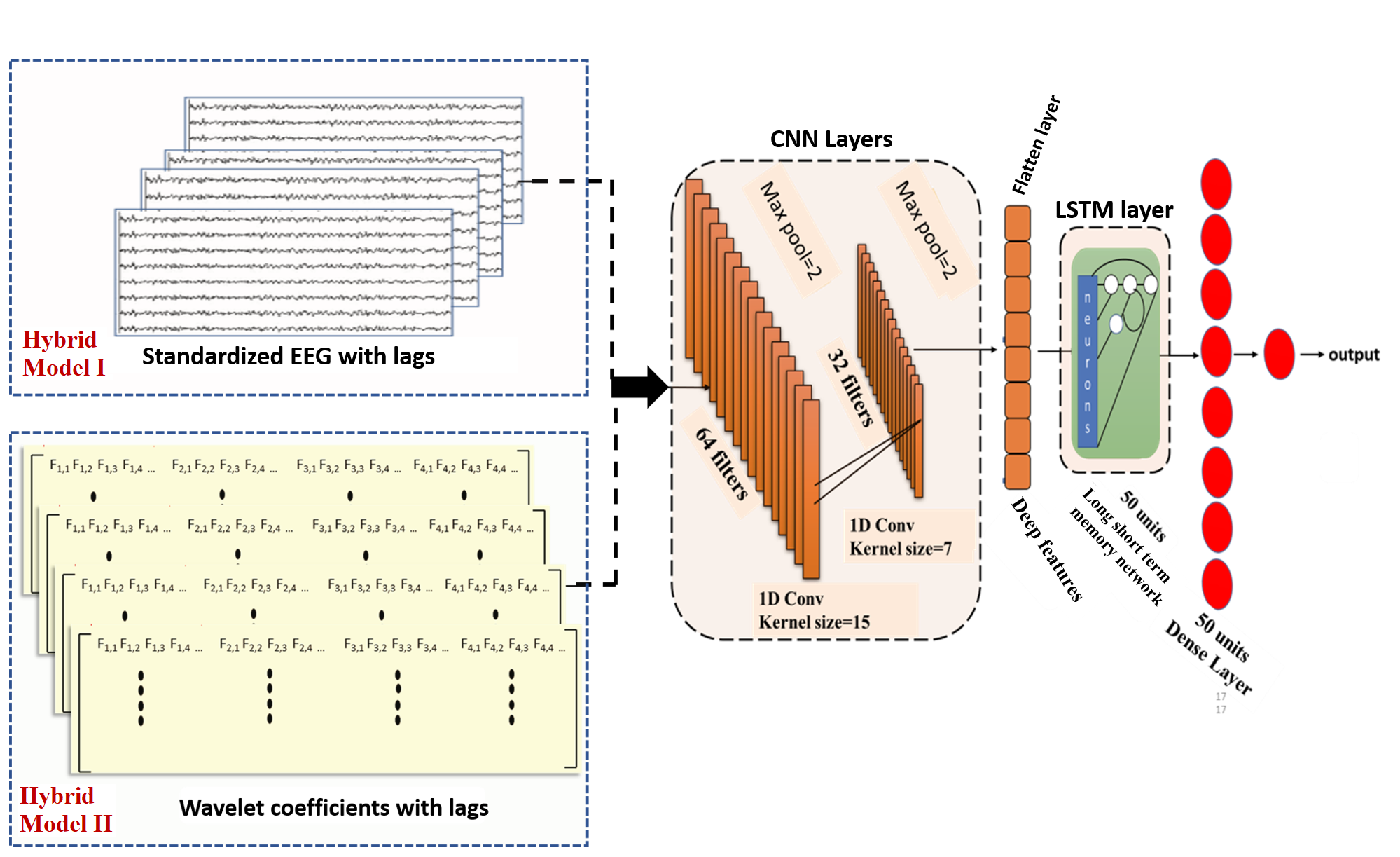}
	\caption{Proposed source aware hybrid model I utilizes EEG time series as an input and hybrid model II takes wavelet coefficients for predicting the kinematics of upper limb.}
	\label{experiment1}
\end{figure*}
 \begin{figure}[t]
\centering
\includegraphics[width=0.5\textwidth]{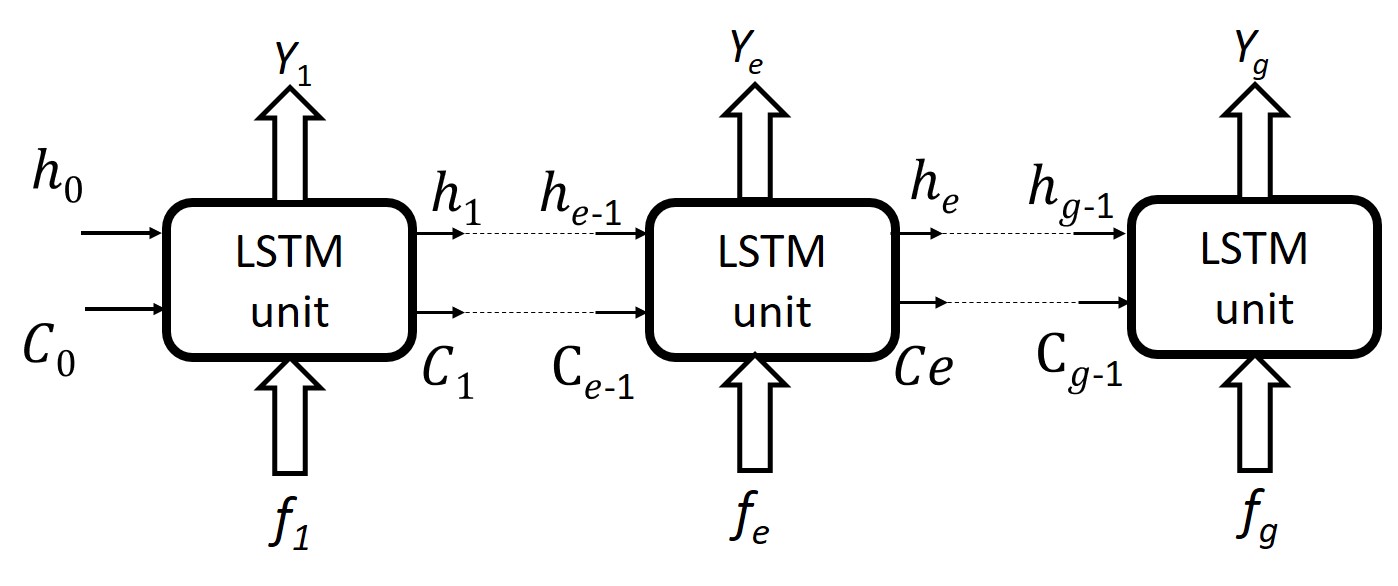}
\caption{Unfold structure of LSTM network.}
\label{LSTM}
\end{figure}
\subsection{Model I : Multi-layer Perceptron Model based}
Deep learning based models are not much explored in the literature for MTP using EEG signal. However, MLP has been utilized for EEG classification \cite{narang2018classification}. Multi-layer perceptron is first proposed herein for trajectory prediction using EEG signal. The model is illustrated in the Fig.\ref{MLP}. The building blocks of a neural networks are neurons (or perceptron), weights and activation functions. For activation function, rectified linear unit (ReLU) is employed herein and is defined as
\begin{align}
F(\Omega)=max(0, \Omega)
\label{relu}
\end{align}
where $\Omega$ is the input parameter to the activation function $F$. The MLP model utilizes a feed-forward neural network consisting of input, hidden and output layers. In particular, there are 3 hidden layers having 300, 100 and 50 perceptrons, in addition to input and output layer. The output of the $d$th neuron at the first hidden layer is given by
\begin{align}
h_d^1 = F(b_d^1 + \sum_{i=1}^{D_0}\ w_{i,d}^0\ V_i), && d=1, \cdots, D_1 
\label{mlp1}
\end{align}
where $b_d^1$ is the input bias, $D_0$ is the size of input vector and $D_1$ represents the total number of neuron at first hidden layer. $w_{i,d}^0$ is the weight between the $i$th element of the input $V_i$ to the $d$th neuron at the first hidden layer. Since, every perceptron in each layer of the neural network is connected to every other perceptron in the adjacent layer, the output of the $d$th neuron at $k$th hidden layer can be expressed as
\begin{align}
h_d^k = F(b_d^k + \sum_{i=1}^{D_{k-1}}\ w_{i,d}^{k-1}\ h_{i}^{k-1}), &&  d=1, \cdots D_k,
\label{mlp2}
\end{align}
where $k=2, \cdots K$ represents the hidden layer with a total of $K$ hidden layers. $w_{i,d}^{k-1}$ is the weight between the $i$th neuron in the $(k-1)$th hidden layer to the $d$th neuron at the $k$th hidden layer. The neural network weights are updated using the Adam optimizer. The output of the $k$th hidden layer can be expressed as
\begin{align}
h^{k} = \begin{bmatrix} h_1^k & h_2^k & \cdots h_{D_k}^{k} \end{bmatrix}^{T}
\label{mlp3}
\end{align}
The output layer $y$ of the model yields the desired kinematic parameters. The $j$th element of the output is given by
\begin{align}
y_j = F(b_{j} + \sum_{i=1}^{D_{K}}\ w_{i,j}^{K}\ h_{i}^{K})
\label{mlp4}
\end{align}
where $b_{j}$ is the output bias. 
\subsection{Hybrid Model I : CNN-LSTM based}\label{sec:num4c}
In this Section, a hybrid deep learning based model is proposed for MTP on the GAL dataset. In particular, CNN and LSTM \cite{hochreiter1997long} based deep learning model along with a dense layer is utilized and is shown in Fig. \ref{experiment1}. Hybrid model I utilizes the pre-processed EEG with time lag $l$ varying from $0$ to $L=10$. This is done to enhance the probability of EEG segment corresponding to the observed kinematic data. It is to be noted that the visual stimulus gets reflected in occipital lobe only after $\sim$80ms. Thus, feeding the delayed EEG segments to the CNN-LSTM model allows the model to learn the weighting  parameters correctly. The CNN algorithm is seen to be useful for feature engineering/extraction through layer-by-layer processing \cite{ma2020dwt}. The proposed model makes use of CNN to extract inherent spatial information present in EEG time series. More specifically, relevant combination of sensors is extracted. 

The proposed CNN architecture consists of two 1D convolutional layers with 64 and 32 filters having kernel size of 15 and 7 respectively. The 1D forward propagation (1D-FP) for the $d$th neuron of the $k$th CNN layer is expressed as
\begin{equation}
Z_{d}^{k}=q_{d}^{k}+\sum_{i=1}^{D_{k-1}}conv1D(J_{i,d}^{k-1},S_{i}^{k-1} )
\end{equation}
where $q_{d}^{k}$ is the bias, $D_{k-1}$ is the total number of neuron at layer $k-1$ and $conv1D$ represents the one dimensional convolution. The $J_{i,d}^{k-1}$ denotes the kernel from the $i$th neuron at layer $k-1$ to the $d$th neuron at layer $k$. The output of $i$th neuron at ${(k-1)}$th layer is represented by $S_{i}^{k-1}$. Intermediate output feature is now given by 
\begin{equation}
y_{d}^{k}=F(Z_{d}^{k})
\end{equation}
 where $y_{d}^{k}$ represents output of $d$th neuron at layer $k$. A max-pooling layer for sub-sampling is employed between the two layers.

Output of the CNN module is fed to the flatten layer for generating the intermediate deep features. The deep features are in turn, input to the LSTM layer. Total 50 cells are used in the LSTM layer for creating enhanced temporal features. The structure of LSTM network for an input feature sequence [$f_1, f_2...f_g$] is illustrated in the Fig. \ref{LSTM}. The hidden state $h_e$ and activation vector $c_e$ at time-step ($e=1,2...g$). The LSTM unit utilizes the past state $h_{e-1}$, $c_{e-1}$ and current state features ($f_e$) to predict the current state output $y_e$. In the whole loop of LSTM, previous information is utilized recursively. The input gate ($i_e$), forget gate ($m_e$) and the output gate ($o_e$) parameters of LSTM are defined as
\begin{align}
i_{e}=\delta (W_{i}\left [h _{e-1},f_{e}+\psi_{i} \right ])
\label{eq1}
\end{align}
\begin{align}
m_{e}=\delta (W_{m}\left [h _{e-1},f_{e}+\psi_{m} \right ])
\label{eq2}
\end{align}
\begin{align}
o_{e}=\delta (W_{0}\left [h _{e-1},f_{e}+\psi_{0} \right ])
\label{eq3}
\end{align} 
where $\delta$ is the logistic sigmoid function, $W$ is the weight matrix and $\psi$ is the bias of each gate. Activation vector and hidden state can now be computed as
\begin{align}
c_{e}=i_{e}\odot tanh(W_{c}\left [h _{e-1},f_{e}+\psi_{c} \right ])+m_{e}\odot c_{e-1}
\label{eq4}
\end{align}
\begin{align}
h_{e}=o_{e}\odot tanh(c_{e})
\label{eq5}
\end{align}
 where, the $\odot$ represents the point wise multiplication. The final output of LSTM layer 
\begin{align}
y_{e}=W_{y}h_{e}+\psi_{y}
\label{eq6}
\end{align}
becomes the input to the Dense layer. The initial state parameter will be derived after model training for subsequent predictions. The output of the dense layer neurons could be given by equation (\ref{mlp1}) with LSTM layer output as the input. The kinematics parameters output can be obtained by the equation (\ref{mlp4}).
\begin{table*}[h]
 \centering
\caption{PCC analysis of (a) mLR, (b) MLP, (c) CNN-LSTM model in different frequency bands. WPD CNN-LSTM correlation coefficient is presented in the end coloumn labeled as WPD.}
    \centering
    \begin{tabular}{|c|c||c|c|c||c|c|c||c|c|c||c|c|c|c|}\hline 
        \multirow{3}{*}{Subject ID} & \multirow{3}{*}{Direction} & \multicolumn{12}{c|}{Frequency Band} & \multirow{3}{*}{WPD}\\ \cline{3-14}
        & & \multicolumn{3}{c|}{Delta (0.5-3 Hz)} & \multicolumn{3}{c|}{Theta (3-7 Hz)} & \multicolumn{3}{c|}{Alpha (7-12 Hz)} & \multicolumn{3}{c|}{Entire (0.5-12 Hz)} &\\  \cline{3-14}
        & & (a) & (b) & (c) & (a) & (b) & (c) & (a) & (b) & (c) & (a) & (b) & (c) &\\ \hline \hline
         \multirow{3}{*}{1} & x &  0.36 & 0.81 & 0.83 & 0.35 & 0.74 &  0.75 & 0.21 & 0.54 & 0.54 &  0.43 & 0.78 & \textbf{0.84} & 0.82\\ \cline{2-15}
          & y &  0.42 & 0.82 & 0.88 & 0.24 & 0.75 &  0.78 & 0.24 & 0.45 & 0.54 &  0.41 & 0.80 & 0.83 & \textbf{0.84}\\ \cline{2-15}
            & z &  0.18 & 0.75 & 0.80 & 0.21 & 0.68 &  0.61 & 0.12 & 0.22 & 0.22 &  0.21 & 0.62 & \textbf{0.81} & \textbf{0.81}\\ \hline \hline
            
             \multirow{3}{*}{3} & x &  0.34 & 0.82 & 0.81 & 0.32 & 0.77 &  0.74 & 0.21 & 0.50 & 0.51 &  0.45 & 0.78 & 0.84 & \textbf{0.85}\\ \cline{2-15}
          & y &  0.41 & 0.80 & 0.86 & 0.20 & 0.75 &  0.73 & 0.25 & 0.42 & 0.50 &  0.41 & 0.80 & \textbf{0.90} & 0.89\\ \cline{2-15}
            & z &  0.17 & 0.77 & 0.80 & 0.22 & 0.66 &  0.62 & 0.11 & 0.21 & 0.21 &  0.23 & 0.63 & 0.82 & \textbf{0.84}\\ \hline \hline
            
             \multirow{3}{*}{4} & x &  0.32 & 0.82 & 0.89 & 0.35 & 0.72 &  0.78 & 0.21 & 0.54 & 0.52 &  0.43 & 0.74 & \textbf{0.85} & \textbf{0.85}\\ \cline{2-15}
          & y &  0.42 & 0.81 & 0.87 & 0.23 & 0.74 &  0.77 & 0.23 & 0.46 & 0.53 &  0.42 & 0.74 & \textbf{0.91} & 0.90\\ \cline{2-15}
            & z &  0.18 & 0.70 & 0.77 & 0.21 & 0.68 &  0.60 & 0.12 & 0.22 & 0.21 &  0.22 & 0.62 & 0.80 & \textbf{0.83}\\ \hline \hline
            
             \multirow{3}{*}{5} & x &  0.33 & 0.79 & 0.83 & 0.35 & 0.75 &  0.76 & 0.22 & 0.54 & 0.53 &  0.44 & 0.79 & \textbf{0.85} & \textbf{0.85}\\ \cline{2-15}
          & y &  0.39 & 0.80 & 0.85 & 0.23 & 0.73 &  0.79 & 0.26 & 0.46 & 0.54 &  0.40 & 0.81 & \textbf{0.90} & 0.89\\ \cline{2-15}
            & z &  0.19 & 0.70 & 0.75 & 0.20 & 0.62 &  0.61 & 0.11 & 0.23 & 0.21 &  0.20 & 0.63 & 0.75 & \textbf{0.78}\\ \hline \hline
        
        \multirow{3}{*}{6} & x &  0.28 & 0.83 & 0.85 & 0.25 & 0.78 &  0.83 & 0.12 & 0.54 & 0.52 &  0.40 & 0.80 & \textbf{0.87} & 0.86\\ \cline{2-15}
          & y &  0.32 & 0.84 & 0.87 & 0.25 & 0.82 &  0.77 & 0.22 & 0.53 & 0.56 &  0.43 & 0.84 & \textbf{0.91} & 0.89\\ \cline{2-15}
            & z &  0.13 & 0.53 & 0.80 & 0.11 & 0.51 &  0.61 & 0.06 & 0.30 & 0.24 &  0.12 & 0.61 & 0.81 & \textbf{0.84}\\ \hline \hline
            
        \multirow{3}{*}{7} & x &  0.32 & 0.71 & 0.83 & 0.23 & 0.56 &  0.80 & 0.23 & 0.54 & 0.57 &  0.38 & 0.77 & 0.83 & \textbf{0.84} \\ \cline{2-15}
          & y &  0.28 & 0.76 & 0.82 & 0.24 & 0.66 &  0.78 & 0.25 & 0.32 & 0.44 &  0.41 & 0.74 & \textbf{0.88} & \textbf{0.88}\\ \cline{2-15}
            & z &  0.29 & 0.53 & 0.78 & 0.31 & 0.53 &  0.62 & 0.09 & 0.15 & 0.26 &  0.21 & 0.58 & 0.76 & \textbf{0.79}\\ \hline \hline
                        
                    \multirow{3}{*}{10} & x &  0.30 & 0.81 & 0.86 & 0.32 & 0.61 &  0.82 & 0.23 & 0.46 & 0.51 &  0.46 & 0.79 & \textbf{0.92} & 0.91\\ \cline{2-15}
          & y &  0.38 & 0.82 & 0.89 & 0.31 & 0.72 &  0.79 & 0.26 & 0.41 & 0.50 &  0.52 & 0.83 & 0.91 & \textbf{0.92}\\ \cline{2-15}
            & z &  0.31 & 0.64 & 0.80 & 0.32 & 0.63 &  0.60 & 0.12 & 0.21 & 0.24 &  0.25 & 0.64 & 0.82 & \textbf{0.86}\\ \hline \hline
            
                    \multirow{3}{*}{11} & x &  0.37 & 0.83 & 0.88 & 0.40 & 0.78 &  0.70 & 0.26 & 0.64 & 0.53 &  0.44 & 0.82 & 0.86 & \textbf{0.87}\\ \cline{2-15}
          & y &  0.43 & 0.83 & 0.92 & 0.41 & 0.80 &  0.78 & 0.28 & 0.52 & 0.59 &  0.43 & 0.84 & \textbf{0.93} & 0.91\\ \cline{2-15}
            & z &  0.17 & 0.74 & 0.78 & 0.15 & 0.64 &  0.65 & 0.12 & 0.25 & 0.23 &  0.20 & 0.71 & 0.82 & \textbf{0.85}\\ \hline \hline
            
                    \multirow{3}{*}{12} & x &  0.48 & 0.81 & 0.87 & 0.36 & 0.75 &  0.77 & 0.26 & 0.53 & 0.49 &  0.46 & 0.84 & 0.87 & \textbf{0.88}\\ \cline{2-15}
          & y &  0.52 & 0.81 & 0.85 & 0.23 & 0.76 &  0.79 & 0.27 & 0.51 & 0.54 &  0.52 & 0.85 & \textbf{0.91} & 0.89\\ \cline{2-15}
            & z &  0.32 & 0.64 & 0.74 & 0.22 & 0.67 &  0.61 & 0.16 & 0.23 & 0.10 &  0.28 & 0.74 & 0.81 & \textbf{0.83}\\ \hline \hline
            
            \multirow{3}{*}{Average} & x &  0.34 & 0.80 & 0.85 & 0.32 & 0.72 & 0.77 & 0.22 & 0.54 & 0.52 & 0.43 & 0.79 & \textbf{0.86} & \textbf{0.86}\\ \cline{2-15}
          & y & 0.40  & 0.81 & 0.87 & 0.26 & 0.75 & 0.77 & 0.25 & 0.45 & 0.53 & 0.44 & 0.80 & \textbf{0.90} & 0.89\\ \cline{2-15}
            & z & 0.21  & 0.67 & 0.78 & 0.22 & 0.62 & 0.61 & 0.11 & 0.22 & 0.21 & 0.21 & 0.64 & 0.80 & \textbf{0.82}\\  \hline  
    \end{tabular}
    \label{tab:tab01}
\end{table*}
\vspace{2em} 
\begin{table*}[t]
\centering
\vspace{0.2cm}
\caption{Comparison with state-of-the-art techniques.}
\begin{tabular}{llllll}
\hline \hline
\multirow{2}{*}{References} & \multirow{2}{*}{Space} &  Variation  & Variation  & Decoding & \multirow{2}{*}{Correlation (mean)}                   \\
& & in Load & in SF & Method & \\\hline
Robinson et. al \cite{robinson2015adaptive} & 2D & N.A. & N.A. &  mLR & (0.60$\pm$0.07) \\ 
Korik et. al. \cite{korik2018decoding} & 3D & N.A. & N.A.&  mLR & \specialcell{x=0.39, y=0.56, z=0.59} \\ 
Sosnik et. al. \cite{sosnik2020reconstruction} & 3D & N.A. & N.A. &  mLR & \specialcell{x=0.25, y=0.50, z=0.48} \\ \hline \hline
Model I  & 3D & Yes & Yes & MLP & \specialcell{x=0.79, y=0.80, z=0.64} \\
\specialcell{Hybrid Model I}  & 3D & Yes & Yes & CNN-LSTM & \specialcell{x=0.86, y=0.90, z=0.80} \\
\specialcell{Hybrid Model II} & 3D & Yes & Yes & WPD CNN-LSTM & \specialcell{x=0.86, y=0.89, z=0.82}  \\ \hline \hline
\end{tabular}
\label{lit}
\scriptsize{\\ \vspace{0.2cm}
Note:  Surface Friction (SF), Not Available (N.A.).}
\end{table*}

\subsection{Hybrid Model II : WPD CNN-LSTM based}
\begin{figure}[t]
\centering
\includegraphics[width=0.4\textwidth]{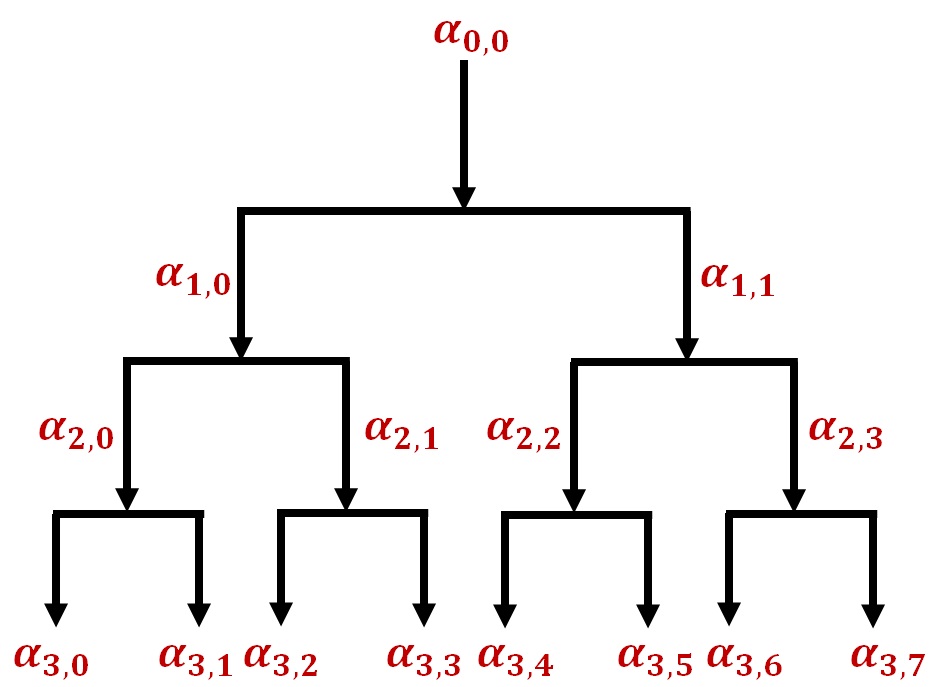}
\caption{Illustration of the three level wavelet packet decomposition.}
\label{wavefig}
\end{figure}

Reconstruction of hand movement profiles using low frequency EEG have been reported in 2D \cite{presacco2011neural} and 3D spaces \cite{bradberry2010reconstructing}. These results indicate that detailed limb kinematic information could be present in the low frequency components of EEG, and could be decoded using the proposed model. Therefore, an advanced version of CNN-LSTM based on wavelet packet decomposition is proposed that decompose the EEG signal into sub-bands with increasing resolution towards the lower frequency band. It may be noted from Fig. \ref{experiment1} that rather than utilising directly the pre-processed time domain EEG signal, wavelet coefficients of the EEG signal are utilized in hybrid model II by employing wavelet packet decomposition \cite{khushaba2010driver,zhang2017classification}. The WPD, also known as optimal sub-band tree structuring, consists of a tree kind of structure with $\alpha_{0,0}$ representing the root node or original signal of the tree as shown in Fig.~\ref{wavefig}. In the generalized node notation $\alpha_{p,r}$, $p$ denotes the scale and $r$ denotes the sub-band index within the scale. The node $\alpha_{p,r}$ can be decomposed into two orthogonal parts: an approximation space $\alpha_{p,r}$ to $\alpha_{p+1,2r}$ and detailed space $\alpha_{p,r}$ to $\alpha_{p+1,2r+1}$. This can be performed by dividing orthogonal basis $\{\theta_p(t-2^p r)\}_{r\in Z}$ of $\alpha_{p,r}$ into two new orthogonal bases $\{\theta_{p+1}(t-2^{p+1} r)\}_{r\in Z}$ of $\alpha_{p+1,2r}$ and $\{\phi_{p+1}(t-2^{p+1} r)\}_{r\in Z}$ of $\alpha_{p+1, 2r+1}$. The scaling function $\theta_{p,r}(t)$ and the wavelet function $\phi_{p,r}(t)$ are defined as
 \begin{align}
\theta_{p,r}(t)=\frac{1}{ \sqrt{\left |2^{p}\right |}}\theta\left ( \frac{t-2^{p}r}{2^p} \right )
 \end{align}
 \begin{align}
\phi_{p,r}(t)=\frac{1}{ \sqrt{\left |2^{p}\right |}}\phi\left ( \frac{t-2^{p}r}{2^p} \right )
 \end{align}
The dilation factor $2^p$, also known as the scaling parameters, measures the degree of compression or scaling. On the other hand, the location parameters $2^pr$ determines the time location of the wavelet function. This method is repeated $P$ times. Total number of samples in the original signal is taken to be $T$, where $P \leq log_2T$. This results in $P \times T$ coefficients. Therefore, at the level of resolution $p$, where $p = 1, 2,...,P$, the tree has $T$ coefficients divided into $2^p$ coefficient blocks or crystals. In this work, Daubechies (db1) is selected as mother wavelet. Total 5 decomposition level are utilized for getting better frequency resolution. WPD coefficients generated at each of the WPD tree subspaces corresponding to approx 120 trials  are utilised as feature matrix. Thereafter, the same deep learning architecture of CNN-LSTM is utilized as discussed in Subsection \ref{sec:num4c}.

To prevent the model from overfitting and underfitting K-fold cross-validation is the most commonly used technique. In the present work, the number of training samples 65000 (120 trials) are divided into three parts : (i) training data for training the model (70\% of the total data), (ii) validation data for hyper-parameter tuning (15\% of the total data), and (iii) test data for performance evaluation (15\% of the total data). To compare the performance of our proposed method (MLP, CNN-LSTM, WPD CNN-LSTM) with existing state-of-the-art techniques (mLR), each model was fed with the same training data. The validation data for all four algorithms were also same. The training and validation data is used to tune the parameters. Once the model was trained, the performance of the model was evaluated on the same test data to ensure that the comparison is fair. Additionally, to make the comparison between the proposed CNN-LSTM and WPD CNN-LSTM model fair, the respective models were built of the same architecture i.e. same number of hidden layers, number of epochs and batch size.
 
\begin{figure*}[!t]
	\centering
	\subfigure[]{\includegraphics[width=0.31\textwidth]{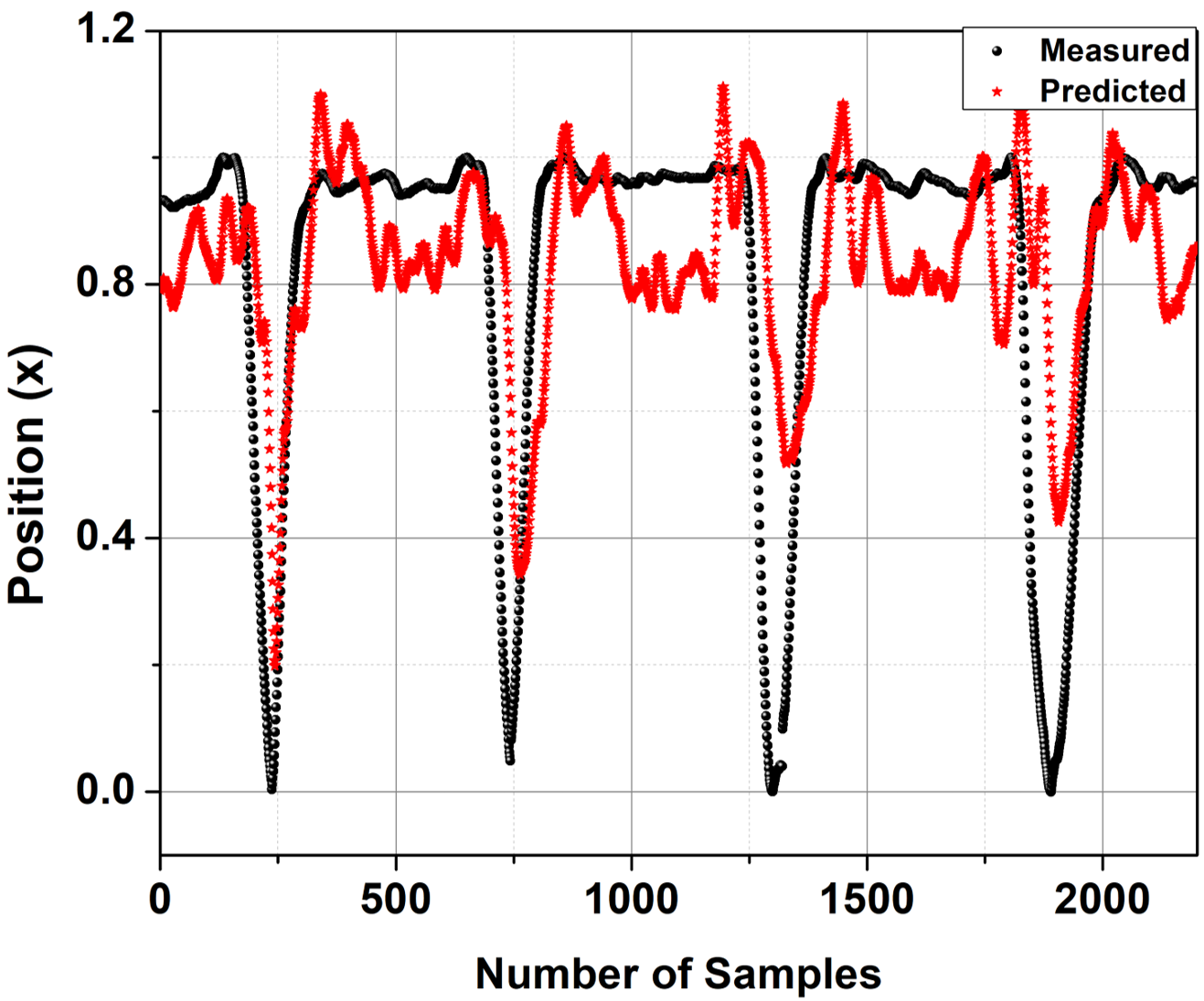}}
	\subfigure[]{\includegraphics[width=0.31\textwidth]{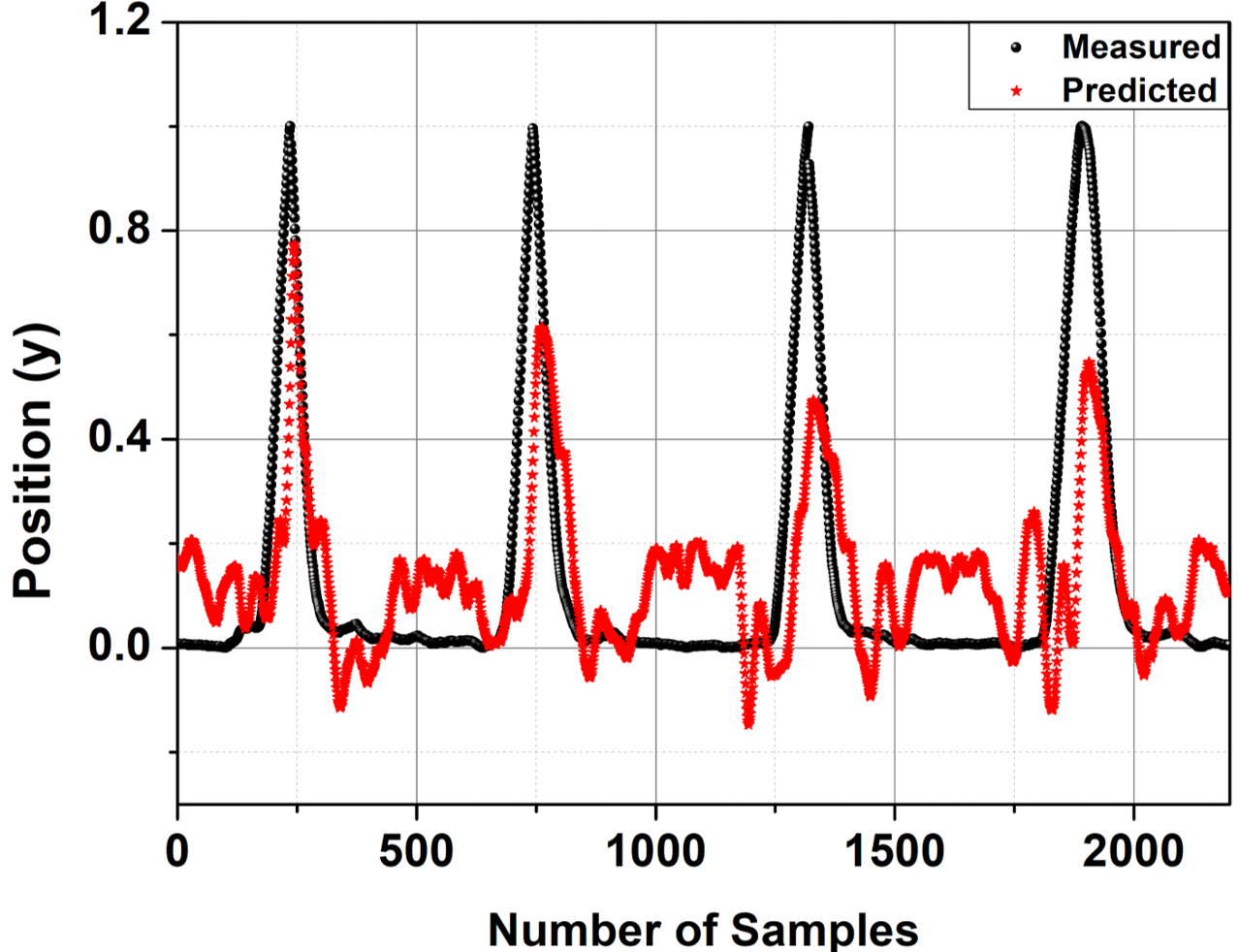}}
	\subfigure[]{\includegraphics[width=0.31\textwidth]{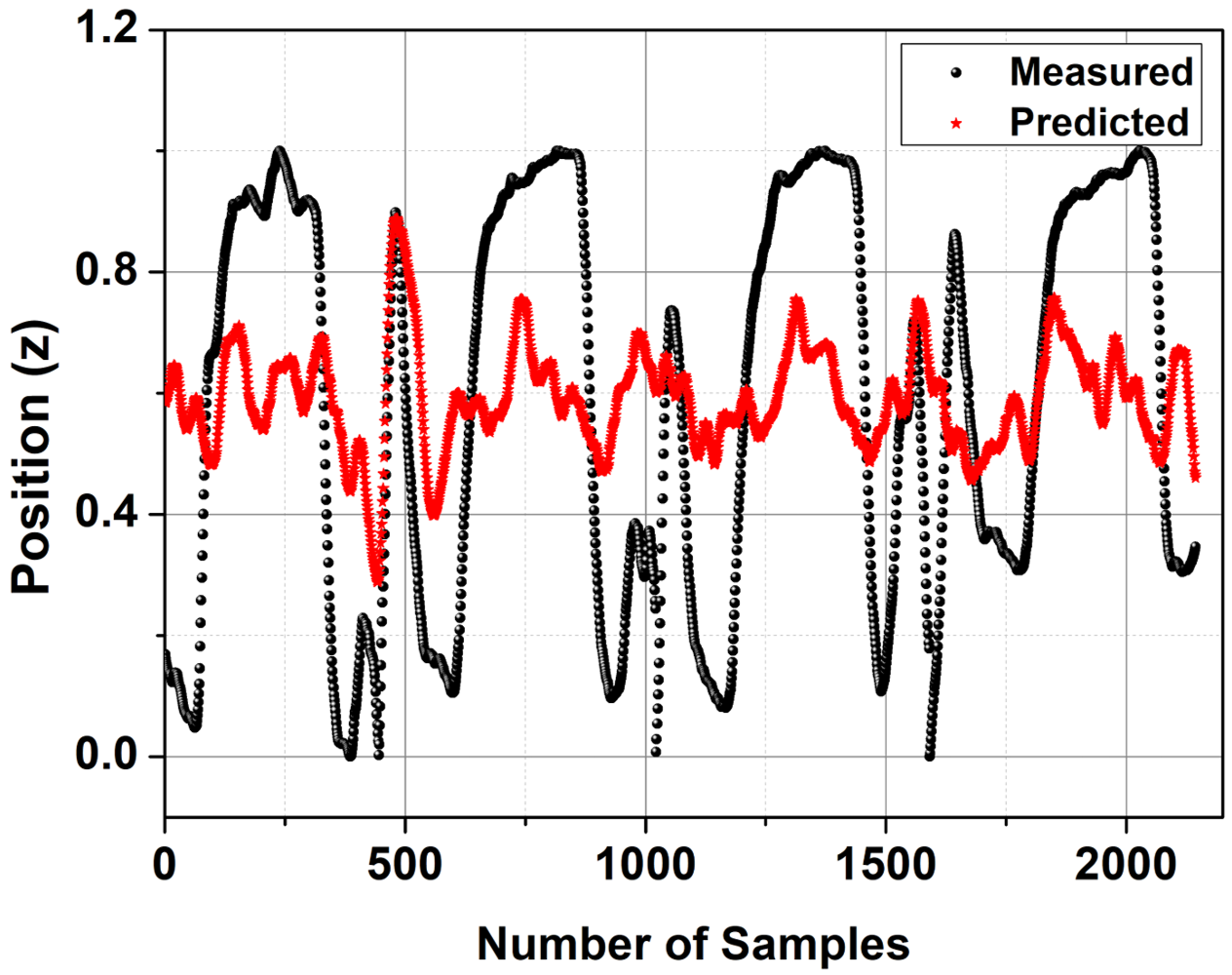}}
	\subfigure[]{\includegraphics[width=0.31\textwidth]{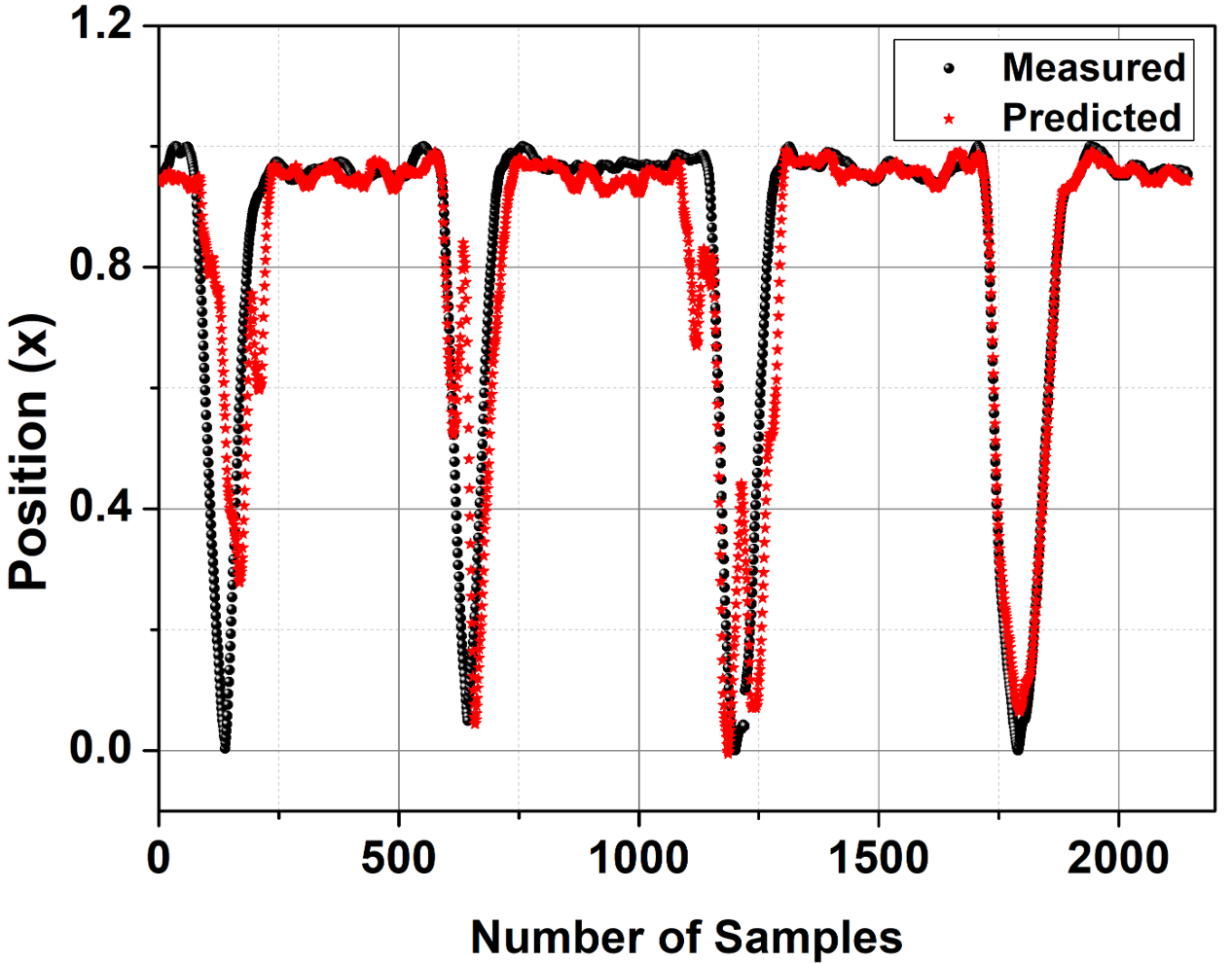}}
	\subfigure[]{\includegraphics[width=0.31\textwidth]{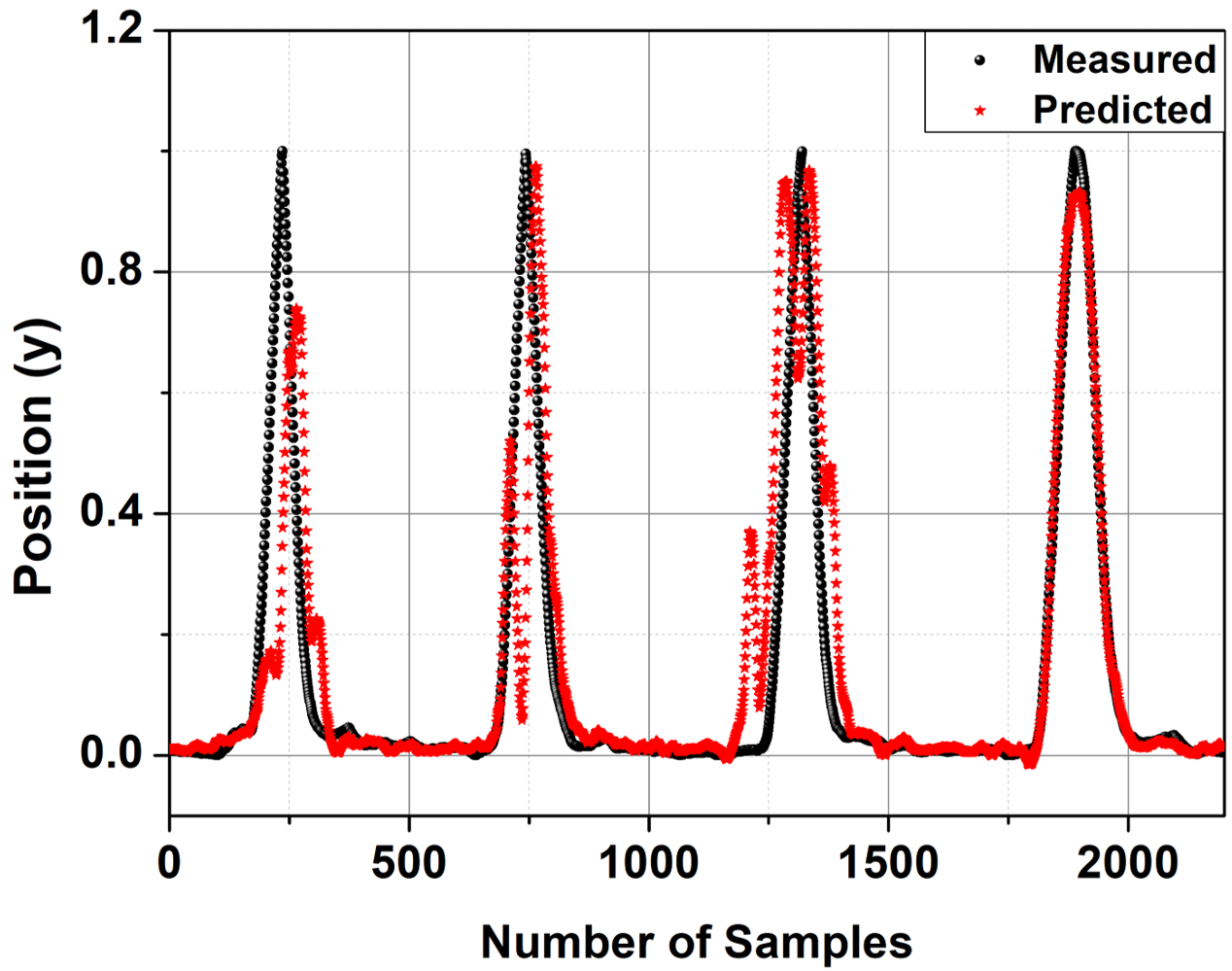}}
	\subfigure[]{\includegraphics[width=0.31\textwidth]{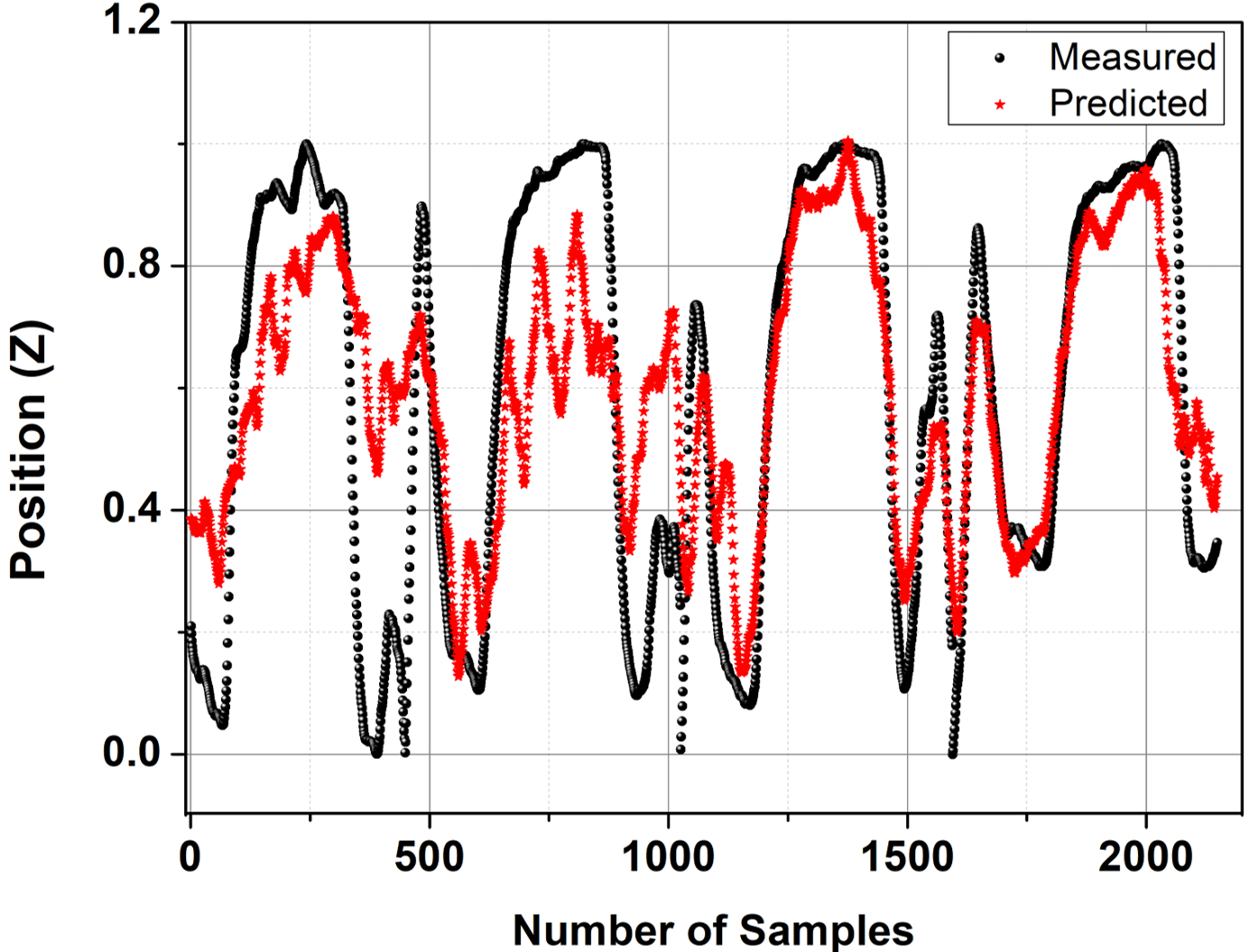}}
	\subfigure[]{\includegraphics[width=0.31\textwidth]{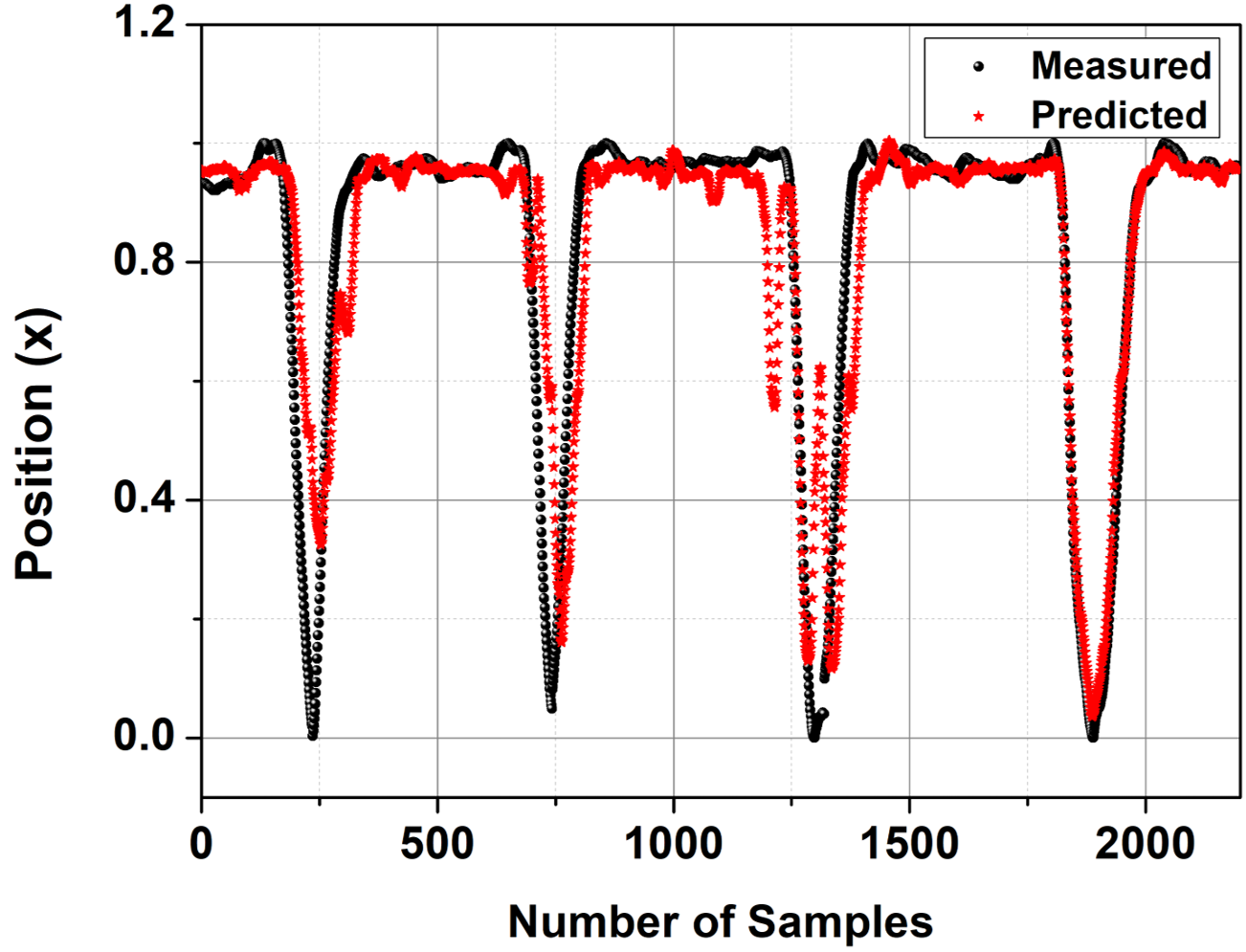}}
	\subfigure[]{\includegraphics[width=0.31\textwidth]{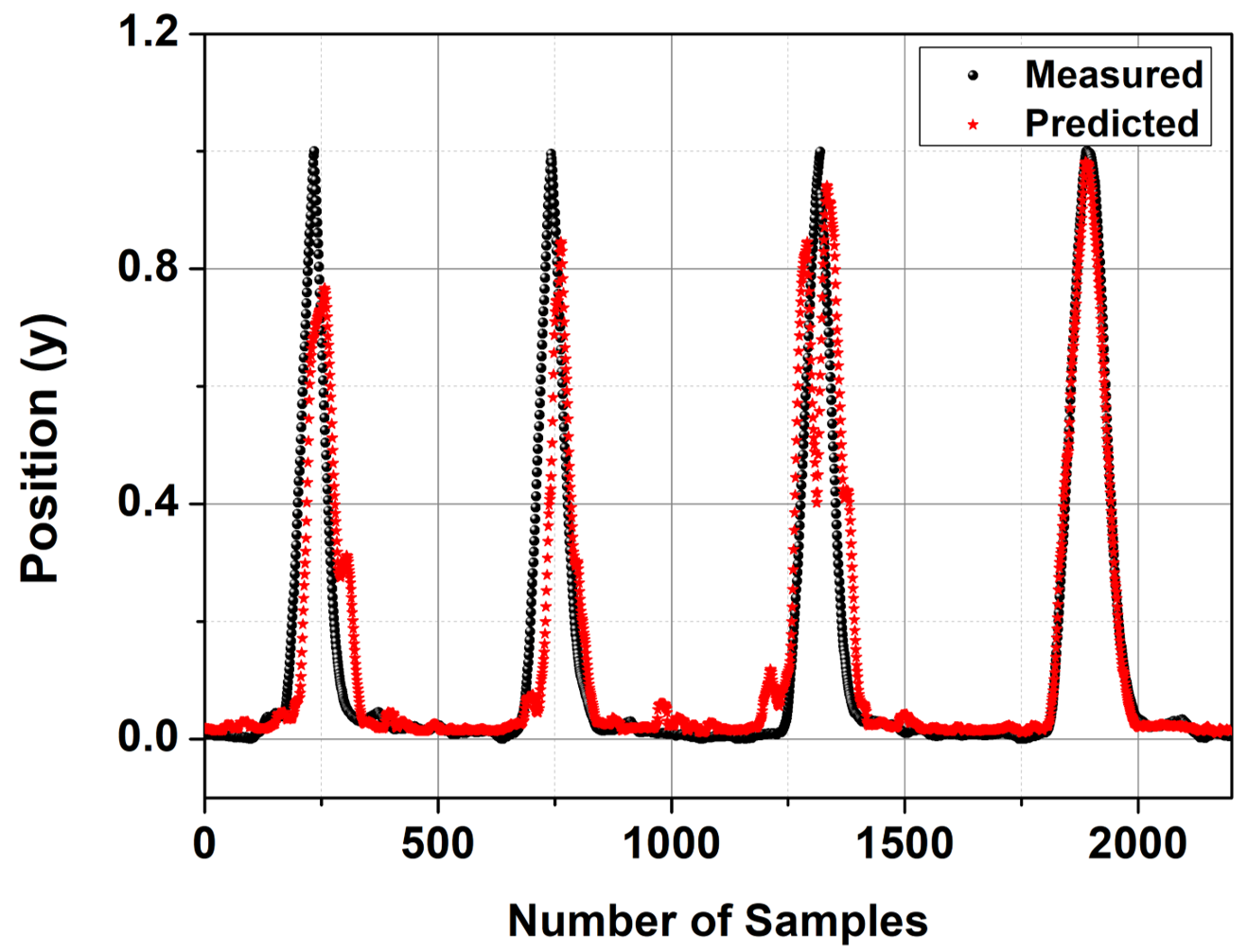}}
	\subfigure[]{\includegraphics[width=0.31\textwidth]{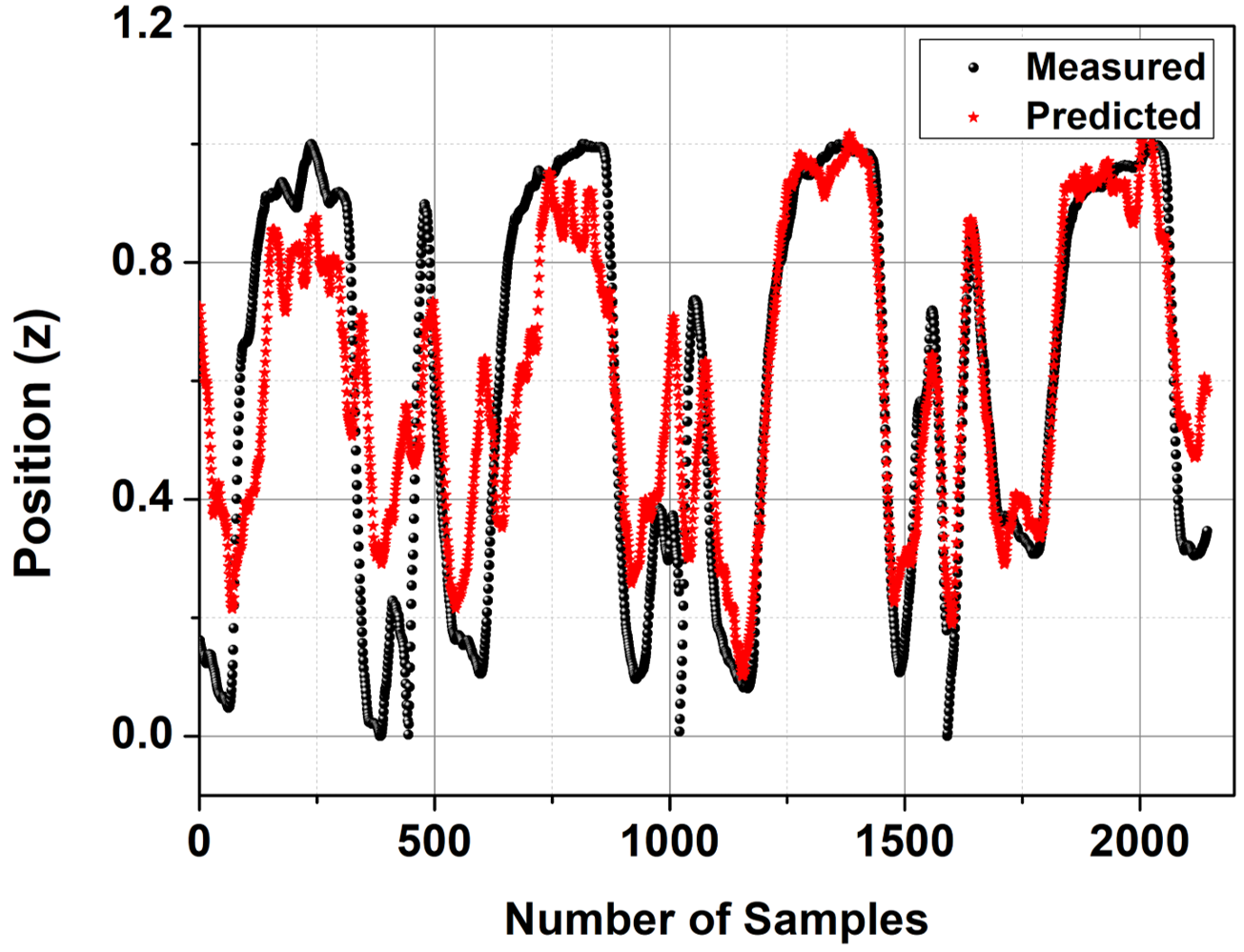}}
	\caption{Trajectory estimation in x, y and z direction using the mLR is presented in (a)-(c) respectively, the CNN-LSTM is presented in (d)-(f) respectively, and using the WPD CNN-LSTM is presented in (g)-(i) respectively.}
	\label{trajectory1}
\end{figure*}
\begin{figure*}[ht!]
	\centering
	\subfigure[]{\includegraphics[width=0.40\textwidth]{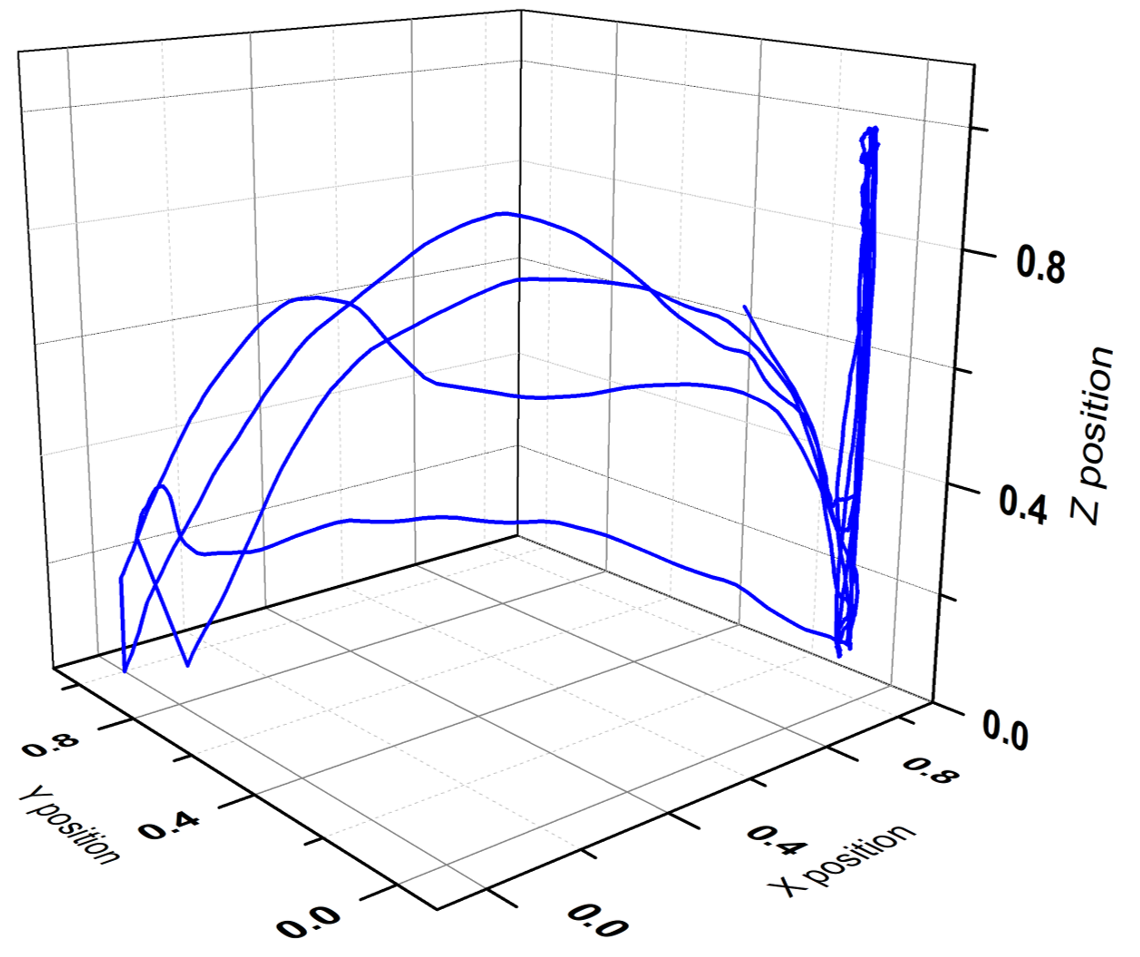}}
	\subfigure[]{\includegraphics[width=0.40\textwidth]{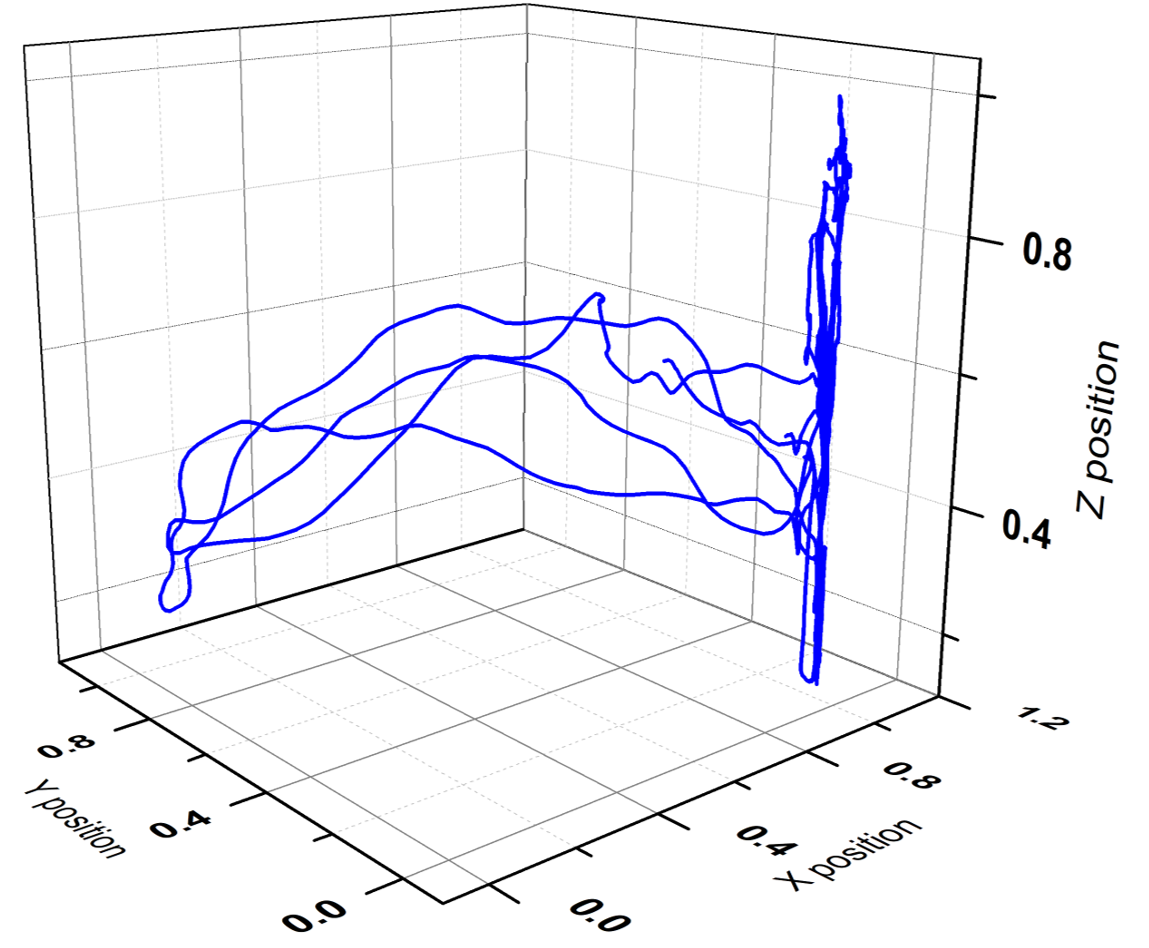}}
\caption{Trajectory visualization in 3D space: (a) ground truth and (b) predicted trajectory from WPD CNN-LSTM method\vspace{0.5cm} }
	\label{trajectory2}
\end{figure*}	
\section{Performance Evaluation} \label{sec:num5}
Performance of the three proposed source aware deep learning models are compared with the state-of-the-art mLR technique. Effectiveness of the proposed framework is established using the Pearson correlation coefficient (PCC) evaluation metric. Additionally, hand trajectory estimation in 3D space is presented and compared with the ground truth trajectory.
\subsection{Pearson Correlation Coefficient Analysis}
Pearson correlation coefficient is a linear correlation coefficient that returns a value between $-1$ to $+1$. A $-1$, $+1$ and $0$ PCC value means there is a strong negative, strong positive and zero correlation respectively. Pearson correlation coefficient between measured ($A$) and estimated ($B$) kinematic parameters of total samples $T$ is expressed as
\begin{align}
	\Pi (A,B)=\frac{1}{T-1}\sum_{i=1}^{T}\left ( \frac{A_i-\upsilon_A}{\sigma_A} \right )\left( \frac{B_i-\upsilon_B}{\sigma_B} \right )
	\label{corr}
\end{align}
where, $\upsilon_x$ and $\sigma_x$ are the mean and standard deviation of $x$ with $x\in \{A,B\}$. The PCC analysis for the various approaches to hand kinematics prediction utilizing EEG signal is presented in Table \ref{tab:tab01} for the 5 subjects from WAY-EEG-GAL dataset. The correlation is presented along the three directions $x$, $y$ and $z$. The time-frequency distribution plot for the dataset in Fig.~\ref{wave} suggests that the EEG frequency power is dominant up to the alpha band. Hence, the PCC analysis is presented in three different frequency bands (Delta, Theta, and Alpha). The correlation was additionally analysed for the entire frequency band. The result is presented for each subject individually along with the average behavior. WPD CNN-LSTM utilises the entire frequency spectrum. The three deep learning based models are compared with the state-of-the-art mLR approach. It may be noted that all the deep learning based models performs reasonably well when compared to mLR method for all the subjects and in all the direction. The low correlation in z direction for all the methods is attributed to short and transit movement along this direction. The performance of the hybrid model I (CNN-LSTM based) is seen to be better when compared to mLR and MLP techniques. The performance of the hybrid model II (WPD CNN-LSTM based) is similar to hybrid model I (CNN-LSTM based) along x and y directions. However, it outperforms all the methods in z direction, achieving higher correlation. Higher correlation in WPD based approach may be attributed to the fact that WPD decomposes the EEG signal into sub-bands with increasing resolution towards lower frequency band. 

Source aware deep learning based models (MLP, CNN-LSTM and WPD CNN-LSTM) have been additionally compared with existing mLR variants \cite{robinson2015adaptive,korik2018decoding,sosnik2020reconstruction}. The comparison is presented in Table \ref{lit}. The three deep learning based models shows superior performance when compared to the existing mLR variants. The utilised WAY-EEG-GAL dataset in the present study, additionally has higher complexity that includes different load variations and surface frictions (SF). 

\subsection{Trajectory Analysis}
Comparative analysis of the true and predicted kinematic trajectories using existing and proposed approaches is presented herein. (x,y,z) coordinate location (or trajectory) of the subject's hand during the task is utilised as kinematic parameter. The measured and predicted trajectories of hand along x,y,z direction are plotted in Fig.\ref{trajectory1} (a)-(c) for mLR, Fig. \ref{trajectory1} (d)-(f) for CNN-LSTM and Fig. \ref{trajectory1} (g)-(i) for WPD CNN-LSTM. In CNN-LSTM model, lower trajectory mismatch is observed along the three x,y,z direction when compared to existing mLR technique. However, the two models suffers with greater trajectory mismatch in z direction. This is overcome in WPD CNN-LSTM model to a greater extent. Reasonably high correlation can be observed for all the three co-ordinates, making it more suitable for real-time application. The hand trajectory visualization of the true and estimated trajectory from WPD CNN-LSTM method in 3D space is additionally, presented in Fig.\ref{trajectory2}. True trajectory (Fig. \ref{trajectory2} (a)) and predicted trajectory (Fig. \ref{trajectory2} (b)) were observed to be nearly equivalent. 

\subsection{Statistical Analysis} To quantify the significance of the three proposed deep learning methods over the existing mLR method, statistical analysis is performed in the entire (0.5-12~Hz) frequency band. We apply the two-sample $t$-test on the two sets of correlation values in $x$, $y$, $z$ direction and overall for each type of proposed method with the existing mLR technique. Statistical test analysis results with $\alpha = 0.05$ are presented in Table \ref{litTT}. It can be observed that the improvements of the three proposed methods are statistically significant as compared to the mLR method. Additionally, the statistical test of the proposed methods among themselves is also presented in Table \ref{litTT}. The improvements in correlation value from CNN-LSTM and WPD CNN-LSTM methods are statistically significant compared to MLP. The improvement in correlation value from WPD CNN-LSTM found not to be statistical significant compared to CNN-LSTM in $x$, $y$ direction and overall. However the results are significant in $z$ direction suggesting better trajectory prediction of WPD CNN-LSTM method in $z$ direction. Overall, the results show that our proposed approach outperforms the existing mLR technique with the best performance observed with the CNN-LSTM and WPD CNN-LSTM method.

\begin{table*}[t]
\centering
\vspace{0.2cm}
\caption{P value of two-sample Statistical $t$-test}
\begin{tabular}{l|l|l|l|l|l|l}
\hline \hline
Direction & A vs B & A vs C  & A vs D  & B vs C &  B vs D & C vs D  \\\hline
x  & $7.49 \times 10^{-15}$ & $2.56 \times 10^{-16}$ & $1.93 \times 10^{-16}$ & $7.57 \times 10^{-05}$ & $6.36 \times 10^{-05}$ & $1$ \\ 
y & $6.90 \times 10^{-12}$ & $2.97 \times 10^{-14}$ & $1.62 \times 10^{-14}$ & $4.76 \times 10^{-05}$ & $5.92 \times 10^{-05}$  & $0.52$\\ 
z & $1.65 \times 10^{-12}$ & $1.87 \times 10^{-16}$ & $1.04 \times 10^{-16}$ &  $3.37 \times 10^{-07}$ & $4.73 \times 10^{-08}$ & $0.05$
\\ \hline  \rule{0pt}{3ex}   
Overall  & $2.11 \times 10^{-19}$ & $1.02 \times 10^{-26}$ & $1.08 \times 10^{-27}$ & $6.79 \times 10^{-7}$ & $5.78 \times 10^{-8}$ & $0.61$ \\ \hline \hline
\end{tabular}
\label{litTT}
\scriptsize{\\ \vspace{0.2cm}
Note:  A - mLR; B - MLP; C - CNN-LSTM; and D - WPD CNN-LSTM}
\end{table*}

\section{Discussion}\label{sec:num6}
\subsection{Decoding Performance of the Proposed Methods}
In the literature, the mLR linear decoder has been widely used to reconstruct movement kinematics. A major limitation of linear decoder is that it demands the linear relationship between the independent (observed event) variables and dependent (predicted event) variable. Human brain is a complex non-linear system. It is therefore more natural to use non-linear methods to analyze the signals generated by such a complex non-linear system. In the present work, three neural networks based methods MLP, CNN-LSTM and WPD CNN-LSTM are proposed. Neural networks infuse non-linearity by adding non-linear activation functions in the hidden and output nodes. Thus the resulting deep learning models can access very descriptive (non-linear) features that define the underlying relationships fairly well and are responsible for the high accuracy in decoding the kinematic parameters. Other than non-linearity, it is also important that the model extracts quality features in spatial, temporal and spectral domain. Generally, the MLP and CNN models work well with the data that has a spatial relationship. Hybrid CNN-LSTM is proposed to extract spatio-temporal quality features from the input data as LSTM has shown excellent performance in extracting temporal information. An advanced version of CNN-LSTM based on wavelet packet decomposition is proposed that decompose the EEG signal into sub-bands with increasing resolution towards the lower frequency band. The WPD CNN-LSTM method thus extracts spatio-temporal and spectral information from the data. This results in superior performance by the proposed three methods compared to the state-of-the-art mLR technique based on their inherent feature extraction capability.
\subsection{Online Trajectory Prediction}
During the training phase (which includes most of the computational burden), the model parameters such as coefficients and weights are estimated. The proposed models have been trained on a system with six core Intel(R) Core(TM) i9-8950HK @2.90GHz CPU and 16 GB RAM. The proposed MLP model took an average execution time of 1.2 seconds per iteration. The average training time of proposed CNN-LSTM and WPD CNN-LSTM model is 19 seconds and 21 seconds per epoch respectively, that is longer than the existing linear mLR models. In addition, the model demands a large amount of input training data to learn its weight parameters. It is to note that the ability of the proposed models to describe a wide range of phenomena and outperform existing state-of-the-art models makes it computationally expensive in the training phase. However, the testing phase of the model in which the hand kinematic trajectory is decoded, is a fast procedure and can be implemented in real time and online BCI system.  To get the predicted trajectory in real time, the model parameters are optimized followed by the translation of model into the embedded system's friendly language e.g. C/ C++. Since, the aim of our work is to lay a foundation of building accurate motion decoders from EEG signals. Implementation of strength augmentation from estimated kinematic parameter is beyond the scope of present work. But, It may be noted that the current framework can be used to predict trajectory estimation for the application of strength augmentation using exoskeleton/ soft exosuit. To date, the majority of available strength augmentation using exoskeleton utilize inertial measurement units (IMU)~\cite{little2019imu,chen2013improvement,beravs2011development}. IMU sensors measure position and velocity which are then combined to reconstruct the trajectory of the movement.

In the current work the output of model is a 3D position coordinates $P_x[t], P_y[t],$ and $P_z[t]$ corresponding to input EEG signals. First derivative of position parameter $P_x[t]-P_x[t-1], P_y[t]-P_y[t-1].$ and $P_z[t]-P_z[t-1]$ represents respectively, the horizontal, vertical and depth velocities of the hand at time sample $t$. The continuous estimation of position and velocity enable strength augmentation using exoskeleton/ soft exosuit in real time.
\subsection{Potential Applications}
In the present work, significance of the proposed models (MLP, CNN-LSTM and WPD CNN-LSTM) in estimating the hand kinematic trajectory is well established. Continuous estimation of hand kinematic trajectories from brain EEG signals has potential in real-time BCI applications for healthy and disabled subjects. The methods for estimating hand position in 3D can provide benefits in stroke rehabilitation therapies (neuro-rehabilitation) and in controlling external arm movement (neuro-prosthetics) to patients with reduced or nonexistent muscle activities due to limb loss or neurological dysfunction. Decoding motor activity directly from brain signals has also attracted attention in muscle power augmentation using neural driven exosuit or exoskeleton devices. This can additionally be utilized by the Army soldiers (healthy subjects) for improved endurance and reduced fatigue.
\subsection{Limitation and Future Work}
The GAL dataset used in our work consists of EEG signals that correspond to the executed hand movement. The four frequency bands : delta (0.5-3 Hz), theta (3-7 Hz), low alpha (7-12 Hz) and entire (0.5-12 Hz) used in the proposed method of the executed movement limit the usage of brain activity information from the marked frequency ranges. It would be interesting to explore the performance of proposed model for the imagined/executed movement in the beta (12-28 HZ) and low gamma (28-40 Hz) bands, in addition to the delta, theta and low alpha band. Another possible extension of the present work is to replace the time-resolved band-pass filtered EEG potential based potential time-series (PTS) input with a time-resolved power spectral density (PSD) based bandpower time-series (BTS).

In the present work, significance of the MLP, CNN-LSTM and WPD CNN-LSTM proposed models in estimating the hand kinematic trajectory is well established. The improvements in correlation value between the true kinematic trajectory and the estimated kinematic trajectory from the three three proposed methods are statistically significant compared to the state-of-the-art mLR method. However the improvement in correlation value from WPD CNN-LSTM method found not to be statistical significant compared to CNN-LSTM in $x$, $y$ direction. But the results are significant in $z$ direction suggesting better trajectory prediction of WPD CNN-LSTM method in $z$ direction. To study the particular characteristic traits of each directional kinematic trajectory from advanced deep learning model WPD CNN-LSTM which results in low correlation in $x$, $y$ direction but not in $z$ direction, requires a large number of trials. A major limitation of the proposed deep learning models is the high computational complexity resulting in long training times. The GAL dataset utilized in our work consists of only 120 trials with 12 subject EEG data out of which 3 Subjects data was excluded due to quality issue. The available trials to optimize the proposed model for low computational cost and to characterize the directional kinematic trajectory are quite less. Therefore, in future, we intend to record large number of trials with large set of subjects. 
\section{Conclusions} 
In this work, source aware deep learning framework is proposed for hand kinematics parameter estimation from non-invasive EEG time series. In particular, MLP, CNN-LSTM and WPD CNN-LSTM models are proposed. An additional novelty of the work is to utilize brain source localization (using sLORETA) for motor intention mapping. The information is utilized for channel selection and accurate EEG time segment selection. Electrodes placed over the active brain region corresponding to the hand movement are utilized, rather than all the available sensors data for efficient computation. It has been observed that the EEG signal can provide the intended hand movement information approximately 350ms prior to actual hand movement. Early detection of intended hand movement is essential in communicating or controlling an external BCI devices. The performance of the proposed models are compared with the state-of-the-art mLR technique on the real GAL dataset. Effectiveness of the proposed framework is established using the Pearson correlation coefficient analysis. Additionally, hand trajectory estimation is presented and compared with the ground truth. Our proposed source aware deep learning models show significant improvement in correlation coefficient when compared with traditionally utilised mLR model. Our current study provides continuous decoding of brain activities that facilitate real time communication between the control block and the actuators block in BCI.

\section*{Acknowledgment}
The authors would like to thank Prof. Shubhendu Bhasin, and Prof. Sushma Santapuri from Indian Institute of Technology Delhi (IITD), and Dr. Suriya Prakash from All India Institute of Medical Sciences (AIIMS) Delhi for their discussion and constructive comments during the preparation of the manuscript.

\balance
\bibliography{Main_Cybernetics}
\bibliographystyle{IEEEtran}

\end{document}